\documentclass[aps,prb,twocolumn,floatfix,footinbib,showpacs,superscriptaddress]{revtex4-1}
\usepackage{graphicx}
\usepackage{amsfonts,amsmath,amssymb,dcolumn}
\usepackage{amsthm}
\usepackage{dsfont,bm}
\usepackage{color}
\usepackage{soul} 
\usepackage{amsbsy}
\usepackage{upgreek}
\usepackage[colorlinks=true,linkcolor=blue,filecolor=blue,menucolor=blue,urlcolor=blue,citecolor=blue,anchorcolor=blue]{hyperref}%
\usepackage{sidecap}
\usepackage{times} 
\usepackage{mathbbol}

\usepackage{sidecap}

\newcommand{\bsigma}{{\boldsymbol \sigma}}
\newcommand{\bnabla}{{\boldsymbol \nabla}}

\usepackage{xspace}
\newcommand{\sns}{$\rm S_1NS_2$\xspace}
\newcommand{\sfs}{$\rm S_1FS_2$\xspace}

\newcommand{\dd}{$\Delta_2/\Delta_1$\xspace}

\newcommand{\La}{\xspace$\Lambda$\xspace}

\begin{document}

\title{Supergap and subgap enhanced currents in asymmetric {$\rm S_1FS_2$\xspace} Josephson junctions}
  
\author{Mohammad Alidoust} 
\affiliation{Department of Physics, Norwegian University of Science
  and Technology, N-7491 Trondheim, Norway}
\author{Klaus Halterman} 
\affiliation{Michelson Lab, Physics Division, Naval Air Warfare Center, China Lake, California 93555, USA}

\begin{abstract}
We have theoretically studied the supercurrent profiles 
 in three-dimensional normal metal and ferromagnetic Josephson configurations, where the magnitude of 
 the superconducting gaps in the
superconducting leads are unequal, i.e., $\Delta_1\neq \Delta_2$, creating asymmetric \sns and \sfs systems. Our results reveal that 
 by increasing the ratio of the 
 superconducting gaps \dd,
 the critical supercurrent in a ballistic \sns system can be
 enhanced by  more than $100\%$,
  and reaches a saturation point, or decays away,
  depending on 
  the junction thickness, magnetization strength, and chemical potential. 
  The total critical current in a diffusive \sns system 
  was found to be enhanced by
   more than $50\%$ parabolically and reaches saturation by increasing one of the superconducting gaps. In a uniform ferromagnetic junction, the supercurrent undergoes reversal by increasing \dd$>1$. 
   Through decomposing the total supercurrent into 
   its supergap and subgap components, 
   our results illustrate their crucial relative contributions to 
   the Josephson current flow.
    It was found that the competition of subgap and supergap currents in a \sfs junction results in the emergence of second harmonics in the current-phase relation. 
    In contrast to a diffusive asymmetric Josephson configuration, 
    the behavior of 
    the
    supercurrent in a ballistic system with \dd$=1$ can be properly 
    described by the
    subgap current component only, in a wide range of parameter sets, including 
    Fermi level mismatch, magnetization strength, and junction thickness. 
    Interestingly, when \dd$>1$, our results have found multiple parameter sets where
    the  total supercurrent is driven by the supergap component. Therefore,  
    our comprehensive study  highlights the importance of subgap and supergap supercurrent components in 
   both the ballistic and diffusive regimes.
   We focus on experimentally accessible material and geometric parameters 
   that can
   lead to  advancements in 
   cryogenic devices based on Josephson junction architectures that utilize  supergap currents, 
   which are less sensitive to temperature compared to the subgap current. 
\end{abstract}

\date{\today}
\maketitle

\section{introduction}

When
 two superconductors with different macroscopic phases
 are weakly coupled by proximity effects,  a finite dissipationless current
 can flow, demonstrating the Josephson effect \cite{B.D.Josephson}. The current flow is carried through the coherent tunneling of Cooper pairs from one superconductor (S)  to the other. The coherent nature of Cooper pairs allows for supercurrent flow through finite-thickness normal (N) metal and ferromagnetic (F) materials (SNS and SFS junctions). 
For these types of junctions, the physical quantities of interest can have complicated variations across the
structure over a wide range of length scales due to proximity induced inhomogeneous superconductivity.

The widely accepted microscopic theory of conventional superconductors is the mean field BCS theory, where two electrons with opposite momenta and spins create a single boson through lattice vibrations. This theory was later reformulated by introducing particle-hole space, which is the well-known  Bogoliubov-de Gennes (BdG) approach \cite{bdg}. To microscopically study systems containing a superconducting segment, one employs the associated BdG Hamiltonian
with pair potential $\Delta(x)$ to account for the spatially varying superconducting  correlations.
The competition between superconducting order and other phases in proximity coupled junctions can
induce striking phenomena that has 
 attracted considerable attention  over the  decades \cite{Z.Radovic,T.Karabassov,S.Acharjee,Z.Shomali2011,
 Hikino,Setiawan,jap_al,s1fs2_ov,Ryazanov1,Ryazanov2,Fominov1,Takahashi,H.Chakraborti,
 T.Karabassov2019,S.V.Bakurskiy,Z.Shomali,M.V.Avdeev,E.Moen2018,K.Ohnishi,
 K.Kulikov,layered,kontos2002,shell2006,zutic,fabian,Mazanik,H.Meng,khold}. Differing approaches and a wide range of approximations have been incorporated to study various normal and ferromagnetic superconducting hybrids that have achieved success to describe experimental observations\cite{D.Culcer,G.Tkachov,A.V.Galaktionov,A.G.Golubov,Iovan1,Iovan2,L.R.Tagirov,M.Alidoust2020,K.Halterman2015,C.-T.Wu2018,E.Koshina,C.W.J.Beenakker,A.A.Golubov,halterman2002}. For instance, a recent study of superconducting (half-)metallic spin-valves has shown good agreement between theoretical predictions and experimental observations\cite{zep,half,bernard1,half2}. 
 
Nevertheless, except in simple situations, it is highly 
challenging to obtain analytical solutions to the BdG Hamiltonian. 
One particular  example is asymmetric Josephson junctions,
 where the pair potentials in the S regions are different, i.e.,   $\Delta_1\neq \Delta_2$.
 There are mainly two approaches for studying  current flow in \sns and \sfs configurations. (i) The wave-function approach where one 
 diagonalizes the BdG Hamiltonian to obtain the wave functions
and energies, which after application of the appropriate boundary conditions, 
permits calculation of  the subgap bound states. 
  To further simplify the resultant expressions, 
  the vast majority of works utilize the
   so-called Andreev approximation. 
   (ii) The other approach is Gorkov's Green function technique \cite{abrikosov}. Here, also one needs to incorporate multiple simplifying assumptions for obtaining simple and solvable equations. The best-known approximation in this approach is the quasiclassical approximation where the Fermi energy is considered the largest energy in the system,
   leading to  the
   Eilenberger equation \cite{eilenberger}. One main advantage of this approach is that is can
    conveniently accommodate 
     nonmagnetic impurities via a white-noise scattering potential. In the presence of disorder and nonmagnetic impurities one can integrate the Eilenberger equation over the random quasiparticle  scattering angle 
     to arrive at the Usadel equation \cite{usadel}. This approach has been recently 
     generalized to a spin-orbit-coupled electron gas to study several phenomena including:
     the spatial distribution of spin currents \cite{alidoust1,alidoust2}, the surface state of three-dimensional topological insulators\cite{zu1,zu3}, Weyl semimetals\cite{AlidoustWS2}, and black phosphorus\cite{AlidoustBP2}.   

The former approach (i) has been used to simulate ballistic systems, where multiple interference effects from the
propagating quasiparticles strongly influences the transport behavior of the system.
 It was demonstrated that this approach, if followed analytically, can be
 problematic for asymmetric \sns junctions even within the quasiclassical regime\cite{bagwell}. The problem becomes increasingly difficult in asymmetric \sfs structures due to the inclusion of band spin splitting. 
 One main issue is to properly obtain the contribution of supergap channels to
 the  total supercurrent. These modes become 
 particularly important in asymmetric junctions \cite{bagwell,A.V.Galaktionov,zu3} 
 due to the imbalance of superconducting gaps that 
 open up the continuum domain to states that can carry  considerable amounts of supercurrent. 
  
In this paper, we aim to study the behavior of the supercurrent in asymmetric three dimensional \sns and \sfs Josephson junctions in both the ballistic and diffusive regimes. Due to the 
asymmetry in the pair  potential ``well'',  
 three  relevant energy scales play a role in the net supercurrent response:
(i)  subgap energies  ($\varepsilon\leq \Delta_1$),  which comprise  the  resonant Andreev bound states,
(ii) supergap  energies ($\Delta_1< \varepsilon\leq\Delta_2$), and
(iii) energies in the continuum, where   $\varepsilon >\Delta_2$. 
 We demonstrate that our microscopic numerical approaches in the ballistic regime can adequately 
 describe the supercurrent flow in all three energy regimes and provides
 an accessible framework that recovers
 previous results in various asymptotic limits for a simpler one-dimensional quasiclassical \sns system \cite{bagwell}. 
 Our numerical approaches allow for exploring realms beyond those studied in the vast majority of the literature without imposing any limitations to Fermi level mismatch and magnetization strength (supporting weak magnetization to a half-metallic phase). 
 Our results reveal that when \dd$=1$, the subgap component of the supercurrent 
  that describes the resonant bound states 
 can properly account for  the
 total supercurrent in a ballistic Josephson junction, regardless of 
 Fermi level mismatch, junction thickness, and magnetization strength. 
 When \dd$>1$,
  we find the the critical current can be strongly enhanced
   in 
 highly asymmetric ballistic \sns junctions. 
 By means of the  current-phase relations, we find that the supergap and subgap supercurrent components can propagate in opposite directions, and within certain regimes, 
 the subgap supercurrent vanishes, so that the total supercurrent arises from supergap states. 
 
 In the full proximity limit of the
 diffusive regime, we find that the
 critical supercurrent can be enhanced by more than $50\%$ when increasing the
 superconducting gap ratio to \dd~$\sim10$, in an asymmetric \sns junction.
 The critical supercurrent also shows an oscillatory behavior in \sfs junctions
 when
  increasing the exchange field intensity,
  and additionally, the supercurrent undergoes  reversals
  as a function of \dd for certain magnetization strengths. 
  By calculating  the
  total current-phase-relation with its subgap and supergap current components, 
  our results reveal that the emergence of 
  a $\sin 2\Delta\varphi$ harmonic close to a current reversal point is the consequence of the intricate
  competition between the  subgap and supergap currents 
  flowing in opposite directions. 
  
The paper is organized as follows. In Sec.~\ref{theory}, we have summarized 
the main equations
which establish  the theoretical framework employed throughout the calculations. In Secs.~\ref{theory-ballistic} and \ref{theory-diff}, we present detailed formulations of the ballistic and diffusive regimes, respectively. In Sec. \ref{results}, the main results and findings are presented. In Secs.~\ref{results-ballistic} and \ref{results-diff}, we discuss the results 
for
 the ballistic and diffusive regimes, respectively. 
 Lastly, in Sec.~\ref{conclusions}, we give concluding remarks.

\section{theory and model}\label{theory}

\begin{figure}[t!] 
\centering
\includegraphics[trim=4.7cm 0.0cm 4.2cm 0.10cm, clip, width=8.0cm,height=6.1cm]{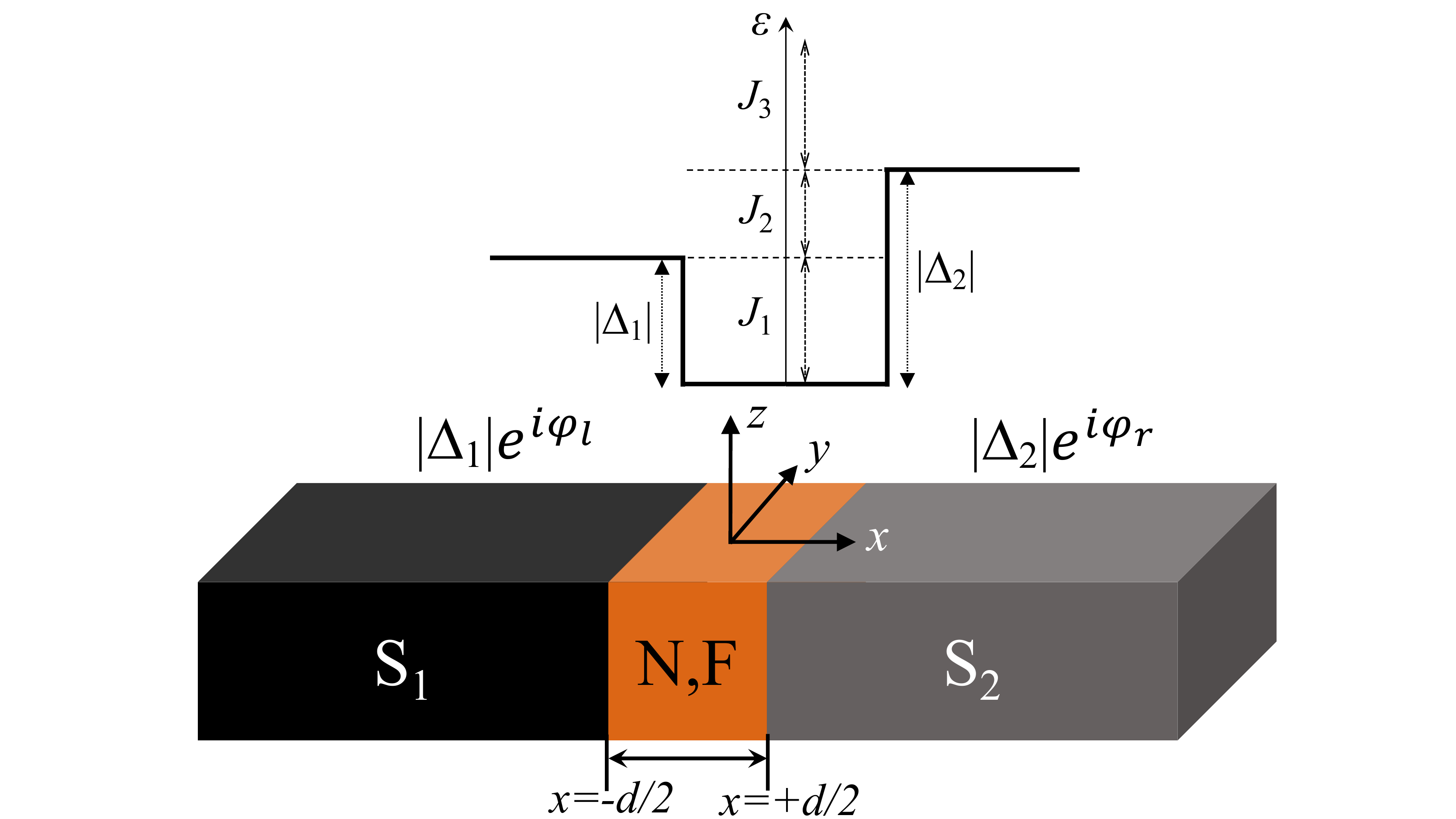} 
\caption{(Color online). Schematic of the asymmetric Josephson junction. 
The left and right superconductors have different superconducting  gaps, $\Delta_{1,2}$, and macroscopic phases, $\varphi_{l,r}$, respectively. The superconductor-nonsuperconductor interfaces are located at $x=\pm d/2$. The two superconductors are connected either by a normal metal (N) or ferromagnet (F), making \sns and \sfs Josephson configurations. The subgap and supergap currents are marked by $J_1$ and $J_{2,3}$, respectively, 
depending on the  quasiparticle energy $\varepsilon$. The superconducting phase difference is defined by $\Delta \varphi=|\varphi_l-\varphi_r|$.}
\label{fig1}
\end{figure}    

In the
ballistic regime, we directly solve the microscopic Bogoliubov-de Gennes (BdG) equations. \cite{bdg} 
In Appendix~\ref{alt} we also outline a complementary numerical method that
can be employed to contrast and compare results.
By employing two 
distinct  numerical 
approaches \cite{halterman2002,K.Halterman2015,C.-T.Wu2018,K.Halterman_ss2016,AlidoustBP1,M.Alidoust2020}. 
it
allows for a 
comprehensive and accurate investigation into general  hybrid Josephson junctions without being limited to 
a narrow  range of ferromagnetic exchange fields  and Fermi level differences.
Moreover, both approaches produce similar results, as expected. 
 In other words, with these methods in the ballistic regime, one is able to span a wide parameter space 
 from weak magnetizations to  half-metallic systems, \cite{half,half2,zep,C.-T.Wu2018} over 
  a
 wide range of Fermi level mismatches between the S electrodes 
 and junction insert. Of course, limiting cases such as  the quasiclassical regime
 can be studied as well.\cite{AlidoustWS1,AlidoustBP1,M.Alidoust2020,AlidoustBP2,AlidoustWS2}
 For systems containing impurities and disorder, we make use of the Usadel equation \cite{usadel} in the full proximity limit of the quasiclassical regime to study the diffusive motion of quasiparticles in asymmetric \sns and \sfs systems. 

\subsection{Ballistic regime}\label{theory-ballistic}

The effective Hamiltonian that describes our asymmetric 
Josephson junction is given by:
\begin{eqnarray}
\label{ham}
{\cal H}_{\rm eff}&=&\int d^3r \left\{ \sum_s
\psi_s^{\dagger}\left(\bm{r}\right) {\cal H}_0
\psi_s\left(\bm{r}\right)\right.\nonumber \\
&+&\left.\frac{1}{2}\left[\sum_{s\:s'}\left(i\sigma_y\right)_{ss'}
\Delta\left(\bm{r}\right)\psi_s^{\dagger}
\left(\bm{r}\right)\psi_{s'}^{\dagger}
\left(\bm{r}\right)+H.c.\right]\right.\nonumber \\
&-&\left.\sum_{s\:s'}\psi_s^{\dagger}
\left(\bm{r}\right)\left(\bm{h}\cdot\bm{\sigma}
\right)_{ss'}\psi_{s'}\left(\bm{r}\right)\right\},
\end{eqnarray}
where $s$,$s'$ are spin indices,
$\bm{\sigma}$ are Pauli matrices, and 
the exchange energy $\bm{h}$
describes the ferromagnet exchange interaction.
The kinetic part of the
single particle Hamiltonian is defined as,
\begin{equation} \label{hoho}
{\cal H}_0(x) = -\frac{1}{2m}
\frac{\partial^2}{\partial x^2}+\varepsilon_\perp -E_F(x),
\end{equation}
in which $\varepsilon_\perp=\frac{1}{2m}(k_y^2+k_z^2)$
 is the quasiparticle
 energy in the $yz$ plane of Fig.~\ref{fig1}. 
 To accommodate the possibility of 
 differing bandwidths in the
 two junction materials,
 we take the Fermi level $E_F(x)$
to equal $E_{FM}$ in the ferromagnet region and $E_{FS}$ in
the superconductor region. 
Following standard procedures\cite{bdg}, we then
utilize
the generalized Bogoliubov transformation, 
$\psi_s=\sum_n\left(u_{ns}\gamma_n
+\eta_sv_{ns}^{\ast}\gamma_n^{\dagger}\right)$,
where  $\eta_s\equiv1(-1)$ for spin-down (up),
to cast Eq.~(\ref{ham})
in terms of
the spin-generalized BdG equations\cite{bdg}:
\begin{widetext}
\begin{align}
\begin{pmatrix}
{\cal H}_0 -h_z&-h_x+ih_y&0&\Delta \\ 
-h_x-ih_y&{\cal H}_0 +h_z&\Delta&0 \\
0&\Delta^*&-({\cal H}_0 -h_z)&-h_x-ih_y \\
\Delta^*&0&-h_x+ih_y&-({\cal H}_0+h_z) \\
\end{pmatrix}
\begin{pmatrix}
u_{n\uparrow}(x)\\u_{n\downarrow}(x)\\v_{n\uparrow}(x)\\v_{n\downarrow}(x)
\end{pmatrix} 
=\epsilon_n
\begin{pmatrix}
u_{n\uparrow}(x)\\u_{n\downarrow}(x)\\v_{n\uparrow}(x)\\v_{n\downarrow}(x)
\end{pmatrix}\label{bogo},
\end{align}
\end{widetext}
where  $u_{ns}$ and $v_{ns}$ are the quasiparticle 
and quasihole amplitudes, respectively (for $s=\uparrow,\downarrow$). 
The  generalized BdG  technique has been  shown to provide a numerically stable framework
for solving  inhomogeneous superconductivity problems \cite{K.Halterman_ss2016,K.Halterman2015,half,half2}.
For the layered Josephson junctions\cite{layered}
considered in this work,
we assume each F and S layer 
is infinite in the $yz$ plane 
and the finite layer thicknesses extend 
along the $x$ axis (see Fig.~\ref{fig1}). 
As a result, the BdG equations are  translationally invariant
in the $yz$ plane, and  become quasi-one-dimensional in $x$. 

 To solve the BdG equation, we first 
 expand \cite{K.Halterman_ss2016,K.Halterman2015} the quasiparticle amplitudes 
in a Fourier series using a complete set of $N$ basis functions:
\begin{align}
\label{basis}
\psi_{n }(x) = \sqrt{\frac{2}{d}}\sum_{q=0}^{N} \sin ({k_q x}) \hat{\psi}_{q} (k_q),
\end{align}
where $\psi_{n} (x)=(u_{n\uparrow}(x),u_{n\downarrow}(x),v_{n\uparrow}(x),v_{n\downarrow}(x))$,
and $\hat{\psi}_{q} =(\hat{u}_{q\uparrow},\hat{u}_{q\downarrow},\hat{v}_{q\uparrow},\hat{v}_{q\downarrow})$.
The wavevector $k_q = q \pi/d$ is discretized by the
thickness of junction $d$. Next, we  transform the real-space BdG equations
by
first inserting  Eq.~(\ref{basis}) into Eq.~(\ref{bogo}) and using orthogonality 
of the basis set to give:
\begin{subequations}
\begin{align}
\label{hok}
\hat{H}_0(q,q')&=\frac{2}{d} \int_{0}^d dx \Biggl\{
\left(\frac{k_q^2}{2m}+\varepsilon_\perp-E_F(x)\right) \nonumber \\
&\times\sin(k_q x) \sin(k_{q'} x)\Biggr\},  
\end{align}
\begin{align}
\label{delk}
\hat{\Delta}({q,q'}) &= \frac{2}{d}\int_{0}^d dx \Delta(x) \sin(k_q x) \sin(k_{q'} x),  
\end{align}
and
\begin{align}
\label{hk}
\hat{h}_{i}({q,q'}) &=\frac{2}{d} \int_{0}^d dx\, h_{i}(x) \sin(k_q x) \sin(k_{q'} x).
\end{align}
\end{subequations}
Here $i=x,y,z$ and we have defined 
$\hat{u}_\sigma=(\hat{u}_{1\sigma},\hat{u}_{2\sigma},\ldots, \hat{u}_{N\sigma})$,
$\hat{v}_\sigma=(\hat{v}_{1\sigma},\hat{v}_{2\sigma},\ldots, \hat{v}_{N\sigma})$.
Additional details on this solution process can be found elsewhere.\cite{khold}

To compute the dc Josephson current, 
we  numerically
diagonalize
the Fourier  transformed BdG equations [with matrix elements in Eqs.~(\ref{hok})-(\ref{hk})]
 to get the eigenenergies $\epsilon_n$ and 
quasiparticle  coefficients $u_{i\sigma}$, $v_{i\sigma}$ ($i=1,...,N$).
The real-space amplitudes are then obtained via the series
expansion in Eq.~(\ref{basis}). Since we wish to determine the current-phase relation for 
asymmetric  Josephson junctions, the input for the pair potential 
is taken to be the bulk gap, $\Delta_1\exp(i\varphi_l)$, in ${S_1}$
and $\Delta_2 \exp(i\varphi_r)$ in $S_2$.
With this form for $\Delta(x)$, 
and
making
 use of the obtained wave functions and eigenenergies, 
 we calculate the charge current with the expression,
\begin{align}
J_x=\frac{2e}{m}\sum_{ns} {\rm Im }\left[u_{ns} \frac{\partial u^{*}_{ns}}{\partial x} f_n+
v_{ns}\frac{\partial v^{*}_{ns}}{\partial x} \left(1-f_n\right) \right],
\label{bdgcurrent}
\end{align}
where $f_n$ is the Fermi function.
The supercurrent satisfies the conservation law
\begin{align}
\label{source}
\frac{\partial J_x (x)}{\partial x} \!=\!
  2e\,{\rm Im}\left\{\Delta({x}) \hspace{-.07cm} \sum_n\left[u_{n \uparrow}^*
v_{n\downarrow}+u_{n\downarrow}^*v_{n\uparrow}\right]
\tanh\left(\frac{\epsilon_n}{2T}\right) \right\}.
\end{align}
Thus, within the junction 
region where $\Delta(x)$ 
vanishes, 
the current density is uniform.
We refer the reader to 
Refs.~\onlinecite{K.Halterman_ss2016,K.Halterman2015} 
for further details on the methods used here for calculating
the supercurrent. Also, an alternative approach to study 
ballistic asymmetric \sns and \sfs systems is described in Appendix \ref{alt}.

\subsection{Diffusive regime}\label{theory-diff}

In a system containing nonmagnetic impurities and disorder, the motion of quasiparticles can be described by a diffusion equation because of the scattering sources. In the quasiclassical  regime, where the Fermi energy is the largest energy scale in the system,  the diffusion equation is given by the so-called Usadel equation \cite{usadel}, 
\begin{subequations}\label{eq:usdl}
\begin{align} 
& D \hat{\bnabla}\{\hat{G}(\varepsilon,\mathbf{r}) \hat{\bnabla} \hat{G}(\varepsilon,\mathbf{r})\} +i  [\varepsilon\hat{\rho}_z, \hat{G}(\varepsilon,\mathbf{r})]=0,\\
& \hat{G}(\varepsilon,\mathbf{r})=\left( \begin{array}{cc}
G^A& G^K \\
0 & G^R
\end{array} \right),
\end{align}
\end{subequations}
where $D$ is the diffusion constant,  $\hat{\bnabla}\equiv (\partial_x,\partial_y,\partial_z)$,
$\mathbf{r}=(x,y,z)$, and the quasiparticle energy
 $\varepsilon$ is measured relative to the Fermi level. The total Green's function 
 $\hat{G}(\varepsilon,\mathbf{r})$ is comprised of the advanced 
$G^A(\varepsilon,\mathbf{r})$, 
retarded $G^R(\varepsilon,\mathbf{r})$, and Keldysh $G^K(\varepsilon,\mathbf{r})$, propagators. 
The $4\times 4$ Pauli matrices in Nambu space are given by $\hat{\rho_z}=\tau_z\sigma_0$ and $\hat{\rho_0}=\tau_0\sigma_0$, in which $\tau_i$ and $\sigma_i$ are $2\times 2$ Pauli matrices in particle-hole and spin spaces, respectively. Throughout the following 
calculations, 
an equilibrium state is considered so that the retarded and Keldysh 
components of the total Green's function can be obtained from the advanced 
component by symmetry considerations. For example, 
 $G^A(\varepsilon,\mathbf{r})=-\{\hat{\rho}_zG^R(\varepsilon,\mathbf{r}) \hat{\rho}_z\}^\dag$ and $G^K(\varepsilon,\mathbf{r})=\{G^R(\varepsilon,\mathbf{r})-G^A(\varepsilon,\mathbf{r})\}\tanh(\varepsilon k_BT/2)$,
 where $k_B$ is the Boltzmann constant,  and the system temperature is denoted by $T$. 
 To simulate the asymmetric Josephson junction shown in Fig. \ref{fig1}, we assume that the 
 superconducting leads are tunnel coupled to the ferromagnetic region and can be 
 described by \cite{boundary_c}:
\begin{eqnarray}\label{eq:bc}
\zeta \hat{G} \mathbf{ n}\cdot\hat{\bnabla} \hat{G} = [\hat{G}, \hat{G}_{\text{S}}],~~ G^R_{\text{S}}=\left( \begin{array}{cc}
{\cal C}& {\cal S} e^{+i\varphi}\\
{\cal S}e^{-i\varphi} & -{\cal C}
\end{array} \right).~~
\end{eqnarray}
Here $\zeta$ is the ratio of the
barrier resistance  to the resistivity of the normal layer, and $\mathbf{ n}$ is the unit vector normal to the interfaces.
The retarded component of the 
total Green's function inside a superconducting lead can be expressed by 
$\hat{G}_{\text{S}}$ so that ${\cal C}\equiv \cosh\theta\sigma_0$, ${\cal S}\equiv i\sinh\theta\sigma_y$,
and  $\theta=\text{atanh}(\Delta/\varepsilon)$. Note that in the case of asymmetric Josephson junction, one should replace $\Delta$ and $\varphi$ by $\Delta_{1,2}$ and $\varphi_{l,r}$, respectively, according to Fig. \ref{fig1}. 

The Usadel equation together with the boundary conditions create a set of coupled complex boundary differential equations. To solve them accurately, we make use of a so-called Riccati parametrization to help in establishing a stable numerical algorithm. 
Two auxiliary unknown 2$\times$2  matrices, i.e., $\gamma$ and $\tilde{\gamma}$, are defined for parameterizing the retarded Green's function:
\begin{eqnarray} \label{eq:gf_riccati}
G^R(\varepsilon,\mathbf{r})=\left( \begin{array}{cc}
(1-\gamma\tilde{\gamma})\Gamma& 2\gamma\tilde{\Gamma} \\
2\tilde{\gamma}\Gamma & (\tilde{\gamma}\gamma-1)\tilde{\Gamma}
\end{array} \right), 
\end{eqnarray}
in which $\Gamma=(1+\gamma\tilde{\gamma})^{-1}$ and $ \tilde{\Gamma}=(1+\tilde{\gamma}\gamma)^{-1}$. 
Implementing the Riccati parameterized Green's function, the Usadel equation, Eq.~(\ref{eq:usdl}), in the nonsuperconducting region of Fig. \ref{fig1} reads
\begin{subequations}\label{eq:usdl_riccati}
\begin{eqnarray}
&&\sum_{k} \Big\{ \partial_{k}^2\gamma-2(\partial_{k}\gamma)\tilde{\gamma}\Gamma\partial_{k}\gamma \Big\}=-2i\frac{\varepsilon}{D}\gamma, \label{eq:usdl_riccati1}\\&&
\sum_{k} \Big\{\partial_{k}^2\tilde{\gamma}-2(\partial_{k}\tilde{\gamma})\gamma\tilde{\Gamma}\partial_{k}\tilde{\gamma}\Big\}=-2i\frac{\varepsilon}{D}\tilde{\gamma}. \label{eq:usdl_riccati2}
\end{eqnarray}
\end{subequations}
Here we have defined $k\equiv x,y,z$ for the spatial 
coordinates.
To account for 
ferromagnetism with an arbitrary 
exchange field, 
i.e., $\mathbf{h}=(h_x,h_y,h_z)$, one simply needs to add $(\mathbf{h}\cdot\mathbf{\sigma})\gamma-\gamma(\mathbf{h}\cdot\mathbf{\sigma}^\ast) $, and $\tilde{\gamma}(\mathbf{h}\cdot\mathbf{\sigma})-(\mathbf{h}\cdot\mathbf{\sigma}^\ast) \tilde{\gamma}$ into the Usadel equation [Eq.~(\ref{eq:usdl_riccati1}) and (\ref{eq:usdl_riccati2}), respectively]. 
Also, the boundary conditions in Eq.~(\ref{eq:bc}) for this parametrization scheme at $x=\mp d/2$ are:
\begin{subequations}\label{eq:bc_riccati}
\begin{eqnarray}
&&\partial_{x}\gamma =\pm(2\frac{{\cal C}_{1,2}}{{\cal S}_{1,2}}+\gamma e^{- i\varphi_{l,r}}-\frac{e^{+ i\varphi_{l,r}}}{\gamma})\frac{{\cal S}_{1,2}\gamma}{\zeta},~~~\\
&&\partial_{x}\tilde{\gamma}=\pm(2\frac{{\cal C}_{1,2}}{{\cal S}_{1,2}}+\tilde{\gamma} e^{+i\varphi_{l,r}}-\frac{e^{-i\varphi_{l,r}}}{\tilde{\gamma}}) \frac{{\cal S}_{1,2}\tilde{\gamma}}{\zeta}.
\end{eqnarray}
\end{subequations}

Finally, the charge current density in the equilibrium state is given by 
 \begin{equation}\label{crt_diff}
 {\bf J}(\mathbf{r}) = \int d \varepsilon \text{Tr}\left\{ \rho_z\left[\hat{G}(\varepsilon,\mathbf{r})\hat{\bnabla} \hat{G}(\varepsilon,\mathbf{r})\right]^K\right\},
 \end{equation}
where `Tr' represents the trace operator. To obtain the total charge current flowing across the junction shown in Fig. \ref{fig1}, one performs an spatial integration over the charge flow component perpendicular to the junction interfaces, namely, $J(x)=\int dy\int dz \;{\bf J}_x(\mathbf{r})$. Due to the charge conservation law, $J(x)$ is a constant within the nonsuperconducting region of Fig. \ref{fig1}. 

\section{results and discussions}\label{results}

In the diffusive regime, all lengths are normalized by the superconducting coherence length in the left superconductor, 
$\xi_S=\sqrt{\hbar D/|\Delta_{1}|}$, and energies are scaled by the superconducting 
gap of the left superconducting electrode at zero temperature, $|\Delta_1|$.
For the ballistic regime,
unless otherwise indicated, 
all lengths are measured in
units of $k^{-1}_F$, 
where $k_F$ is the Fermi wavevector in the S regions.
We also have the 
 dimensionless zero-temperature  coherence length
 $k_F \xi_0 = (2/\pi)(E_{FS}/\Delta_1)$,
 and fix $k_F \xi_0 = 100$.
As in the diffusive regime, energies are normalized by $\Delta_1$.
The measure of mismatch between the Fermi levels is
given by the ratio
$\Lambda = E_{FM}/E_{FS}$.\cite{halterman2002,halterman2004,halterman2005}
Throughout this paper, we assume $\Delta_2>\Delta_1$,
and  we consider a uniform magnetization oriented along the $z$ direction so that $\mathbf{h}=h_z=h$. 
Dimensionless units are implied with
 $\hbar=k_B=1$.

To gain a detailed view
of  the  supercurrent profile in asymmetric 
Josephson junctions, we divide the supercurrent into its three constituent parts.
According to Fig.~\ref{fig1}, these parts consist of 
(i) the subgap supercurrent $J_1$, with energies less than $\Delta_1$,  
(ii) the supergap supercurrent $J_2$, with energies $\Delta_1< \varepsilon<\Delta_2$, and
(iii) the supergap supercurrent $J_3$, for scattering states with energies larger than $\Delta_2$. 
When determining the supercurrent via 
Eq.~(\ref{bdgcurrent}) or Eq.~(\ref{crt_diff}),
 we divide the energy integrals into three parts: 
 $J_1=\int_0^{\Delta_1}\square ~d\varepsilon$, $J_2=\int_{\Delta_1}^{\Delta_2}\square ~d\varepsilon$, and $J_3=\int_{\Delta_2}^\infty \square ~d\varepsilon$. For all approaches,
 the critical supercurrent is calculated 
 in the usual way by finding the maximum of 
 the supercurrent within a phase difference interval of 
 $\Delta\varphi\in [0,2\pi]$,
 namely, $J_c=\max[J(\Delta\varphi)]$.

\subsection{Ballistic \sns and \sfs Josephson junctions}\label{results-ballistic}
 
To study the supercurrent profile  in ballistic Josephson configurations with asymmetric superconducting gaps, 
we employ the quantum particle-in-a-box formalism described above \cite{K.Halterman_ss2016}. 
We emphasize that
 the alternative approach, described
  in Appendix~\ref{alt}, produces 
  similar results and has been used to study various systems, including Rashba-Dresselhaus spin-orbit coupled, type-II Weyl semimetals, and black phosphorus\cite{AlidoustWS1,AlidoustBP1,M.Alidoust2020,AlidoustBP2,AlidoustWS2}. 
Our comprehensive numerical treatment was also found to agree
with certain asymptotic limits, including 
previous results\cite{bagwell} that considered
quasiclassical one-dimensional \sns junctions within the short and long junction limits. 
These limitations were mainly imposed so that
analytical  solutions could be found. 
Our fully microscopic approach 
however does not suffer from these limitations and allows for
investigations into asymmetric Josephson configurations with more
 complicated 
 band structures \cite{K.Halterman_ss2016,K.Halterman2015,M.Alidoust2020,AlidoustBP1,AlidoustWS1,AlidoustBP2}.
  In what follows, our results 
  cover a broad range of geometrical and material parameters, 
  including junction thickness, Fermi energies, superconducting gap ratio, and magnetization strength.
  With regards to normalization schemes, 
   the supercurrent density $J$ is normalized by 
    $J_0\equiv e n_e v_F$, where $n_e$  is
  the bulk electron density and $v_F$ is the Fermi velocity.
For clarity, 
 plots involving the supercurrent  are also scaled by $10^{-2}$. 

\begin{figure}[t!] 
\centering
\resizebox{0.49\textwidth}{!}{
\includegraphics{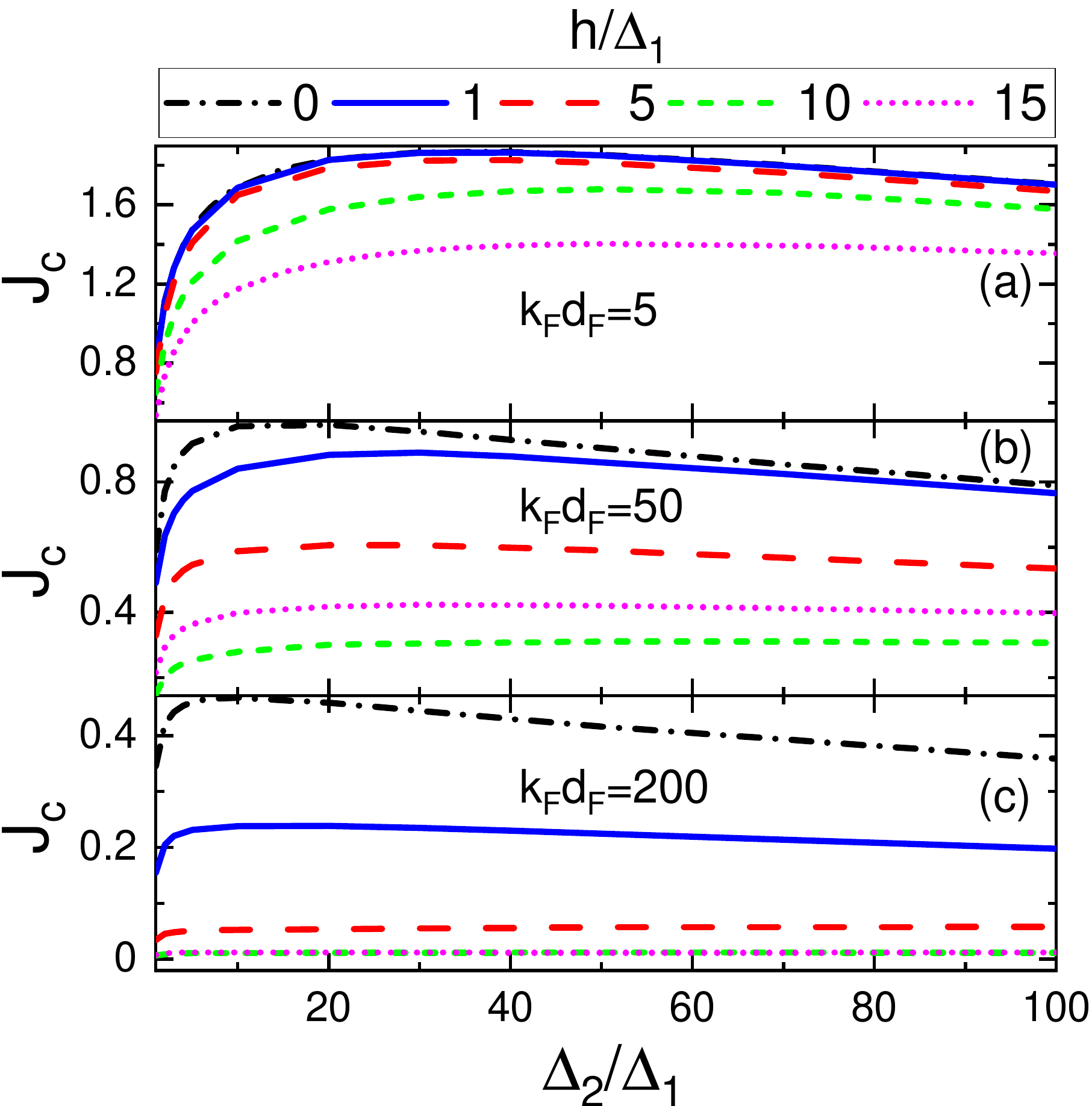}}
\caption{(Color online). The critical supercurrent $J_c$ as a function of superconducting gap ratio $\Delta_2/\Delta_1$ for three values of the  normalized junction thickness: $k_Fd_F=5, 50, 200$, and five values of magnetization strength $h/\Delta_1=0,1,5,10,15$. There is no Fermi level mismatch ($\Lambda=1$).}
\label{fig2}
\end{figure}

To begin, we plot the
critical supercurrent as a function of \dd in Fig. \ref{fig2}. The strength of 
the  magnetization in the central junction region [see Fig.~\ref{fig1}]
varies as $h/\Delta_1=0, 1,5,10,15$,
corresponding to both \sns and \sfs systems.
Each panel in  Figs.~\ref{fig2}(a)-\ref{fig2}(c) examines a different
junction thickness, with $k_Fd_F=5, 50, 200$, respectively. 
Considering the nonmagnetic cases first ($h=0$), 
it is seen that
 the correlation between the two superconducting leads (and thus the supercurrent),
  decays by increasing the junction thickness. 
This effect becomes more pronounced when now considering uniformly magnetized ferromagnets, 
as the pair-breaking exchange field in the magnet tends to 
induce damped oscillations in the Cooper pair wave function with a characteristic decay
that goes as $1/h$.
This causes $J_c$ to become vanishingly small for
 $k_Fd_F=200$ and $h/\Delta_1>5$.
 Note that the results in Fig.~\ref{fig2} have no Fermi level mismatch ($\Lambda=1$),
 which when present can
 amplify the supercurrent significantly,  as will be seen below.
One pronounced feature seen in Fig.~\ref{fig2}(a)
for the short junction limit (i.e., $k_Fd_F=5$),
 is the enhancement of the critical supercurrent 
by more than $100\%$ when 
 increasing the gap ratio to \dd$\approx 25$,
for $h/\Delta_1<5$. 
This enhancement is diminished as the junction thickness increases whereas 
the maximum enhancement of 
the critical current occurs at 
significantly  lower ratios of \dd$\approx 1$.  
Also, as $h$ increases, the critical current maximum
gets shifted to larger $\Delta_2/\Delta_1$ ratios and becomes relatively insensitive to
changes in gap asymmetry beyond $\Delta_2/\Delta_1 \approx 10$.

\begin{figure}[t!]
\centering
\resizebox{0.475\textwidth}{!}{
\includegraphics{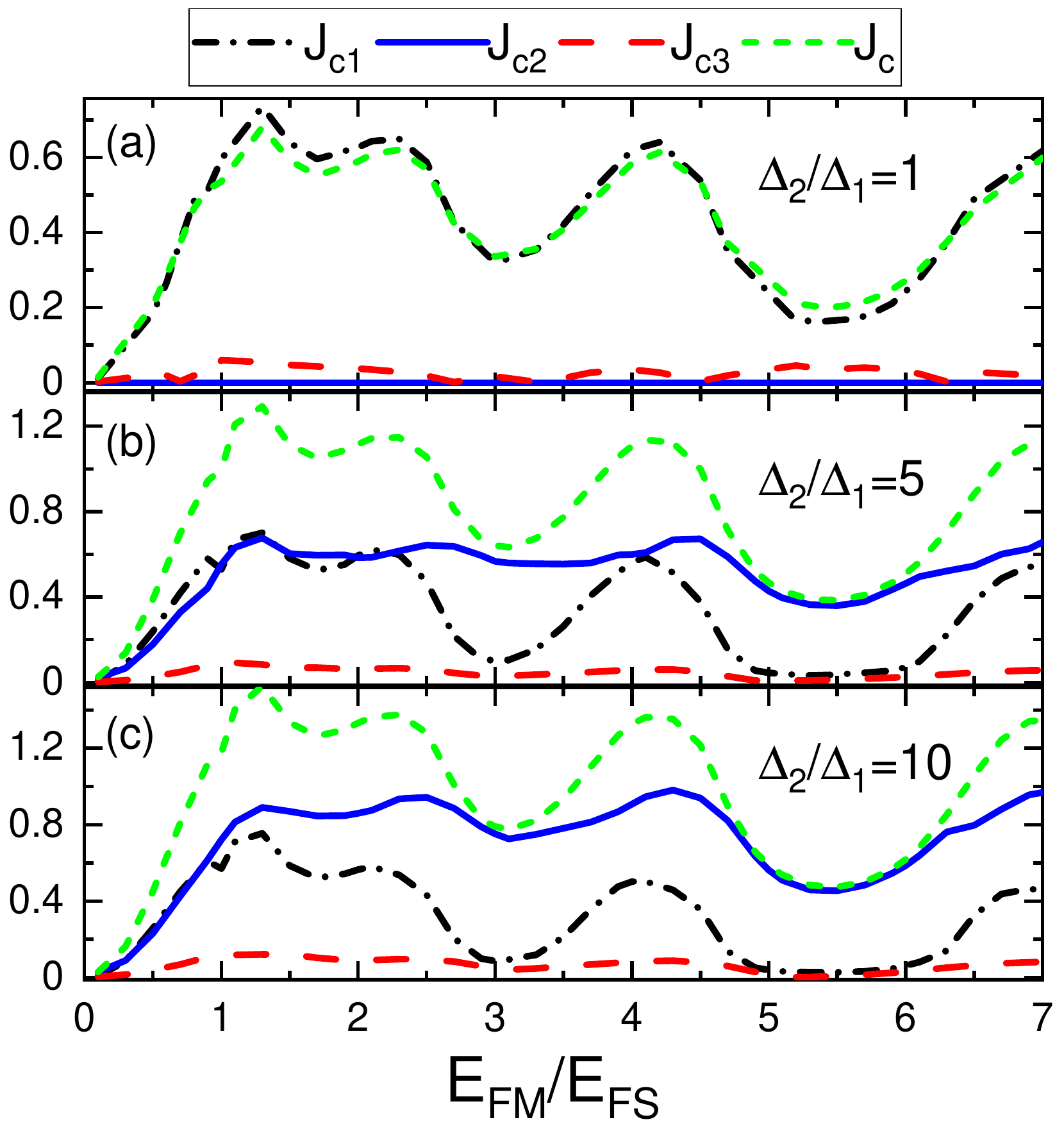}}
\caption{(Color online). Total critical supercurrent  
with its supergap $J_{c2,c3}$ and subgap $J_{c1}$ components as a function of 
the Fermi level mismatch ratio \La. The labels show the different parameter values used for the calculations. The junction thickness is set to $k_Fd_F=5$ and the ferromagnetic exchange field is $h/\Delta_1=15$.}
\label{fig5} 
\end{figure}

We next investigate the effects of Fermi level mismatch, characterized  
by the ratio of the Fermi levels in the
two regions: $\Lambda=E_{FM}/E_{FS}$. For F/S junctions,
 it was previously found \cite{gap,halterman2002,halterman2004,halterman2005} that
the characteristic damped oscillations  of the singlet  pair correlations within the ferromagnet
become drastically modified for $\Lambda<1$.
When there is 
Fermi level mismatch
between the junction layers, the 
energy gap of the system \cite{gap} tends to close,
as revealed in signatures of the density of states \cite{gap,halterman2004,halterman2005}.
In Fig. \ref{fig5}, we plot total critical supercurrent and its subgap 
and supergap components against the Fermi energy 
ratio \La for a \sfs junction with $h/\Delta_1=15$.
Two cases of gap anisotropy are shown in Fig.~\ref{fig5}:
  (b) \dd$=5$ and  (c) \dd$=10$.
The symmetric case (\dd$=1$) is also shown in panel (a)  for comparison purposes.
The subgap critical current ($J_{c1}$) and 
 the
supergap ($J_{c2},J_{c3}$) critical currents 
are  defined as  $J_{ci} \equiv |J_i(\Delta\varphi_c)|$,
where $\Delta\varphi_c$ is the phase angle that leads to largest magnitude of
the total supercurrent, i.e., $|J(\Delta\varphi_c)| = J_c$.
The results reveal that the  total 
critical supercurrent oscillates as a function of \La. The overall profile of $J_c$
remains approximately the same  for all gap ratios \dd, with 
an overall amplification of the magnitude of the supercurrent when 
gap anisotropy is present.
For the symmetric 
 case \dd$=1$  in Fig. \ref{fig5}(a), the $J_{c2}$ supergap current component 
vanishes and $J_{c3}$ is negligibly small.
Thus, nearly the entire contribution to the critical supercurrent comes from 
the $J_{c1}$ subgap 
component.  This is a general feature that
arises in short Josephson junctions with symmetric gaps,
and it is clear that the Andreev bound states are the dominate mechanism for 
supercurrent flow.
Increasing the 
superconducting gap ratio to \dd$=5$, 
opens up the transport channel for states between $\Delta_1$ and $\Delta_2$,
and thus the $J_{c2}$ supergap current component 
can contribute considerably to the total supercurrent. 
Increasing the asymmetry  further to \dd$=10$, 
Fig.~\ref{fig5}(c) shows
that the $J_{c2}$ contribution becomes even greater.
In general, as $\Lambda\rightarrow 0$, the number of available states 
for supercurrent flow declines to zero. This also is true for
the other  extreme case of mismatch
with $J_c =0$  as $\Lambda \rightarrow\infty$.
For the continuum of states 
with energies exceeding $\Delta_2$,
the quasiparticles 
 are no longer confined to the 
gap regions and lose phase-coherence,
resulting in $J_{c3}$
being small relative the other supercurrent components.

\begin{figure*}[t!] 
\centering
{\includegraphics[width=0.49\textwidth]{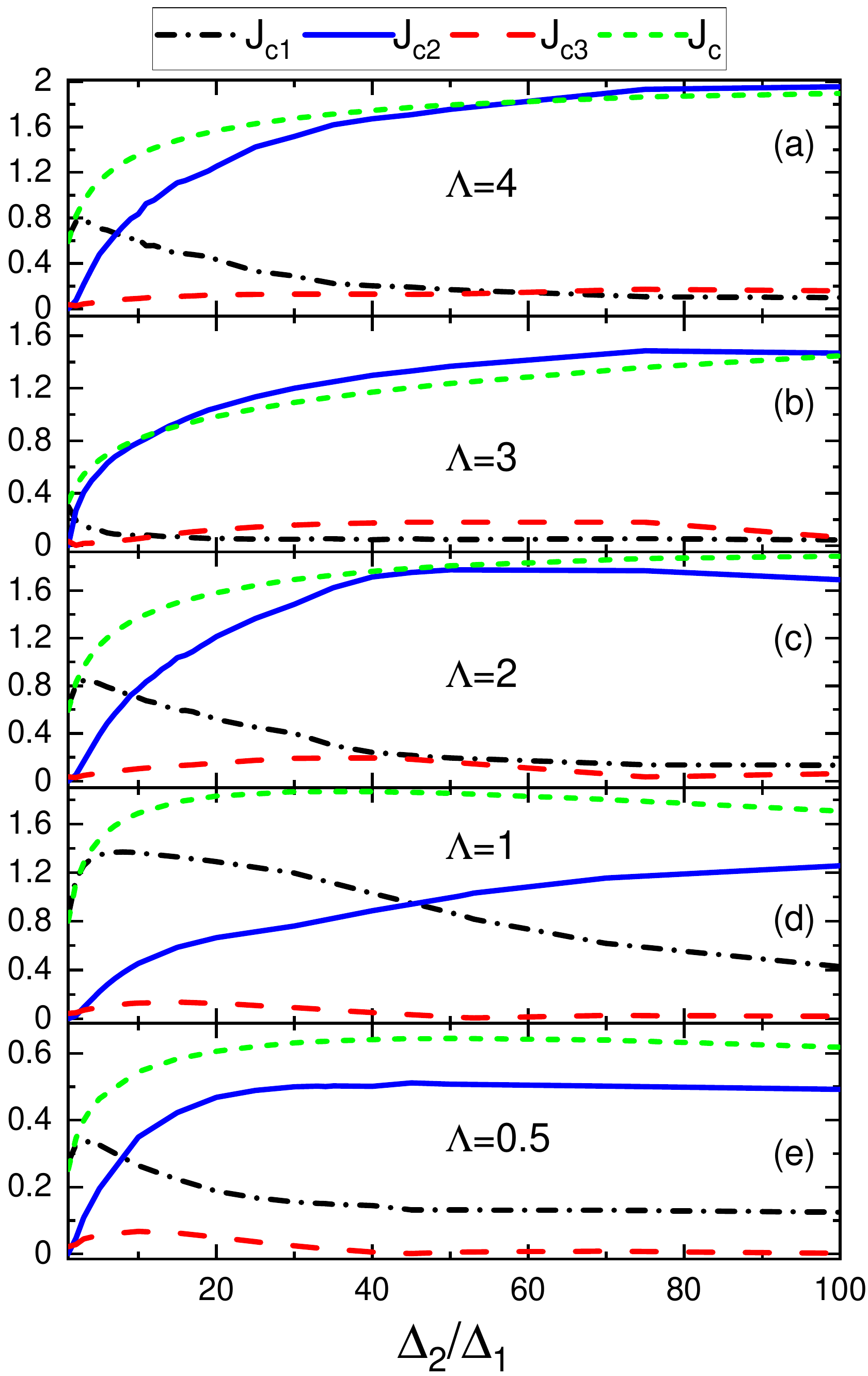} 
\includegraphics[width=0.49\textwidth]{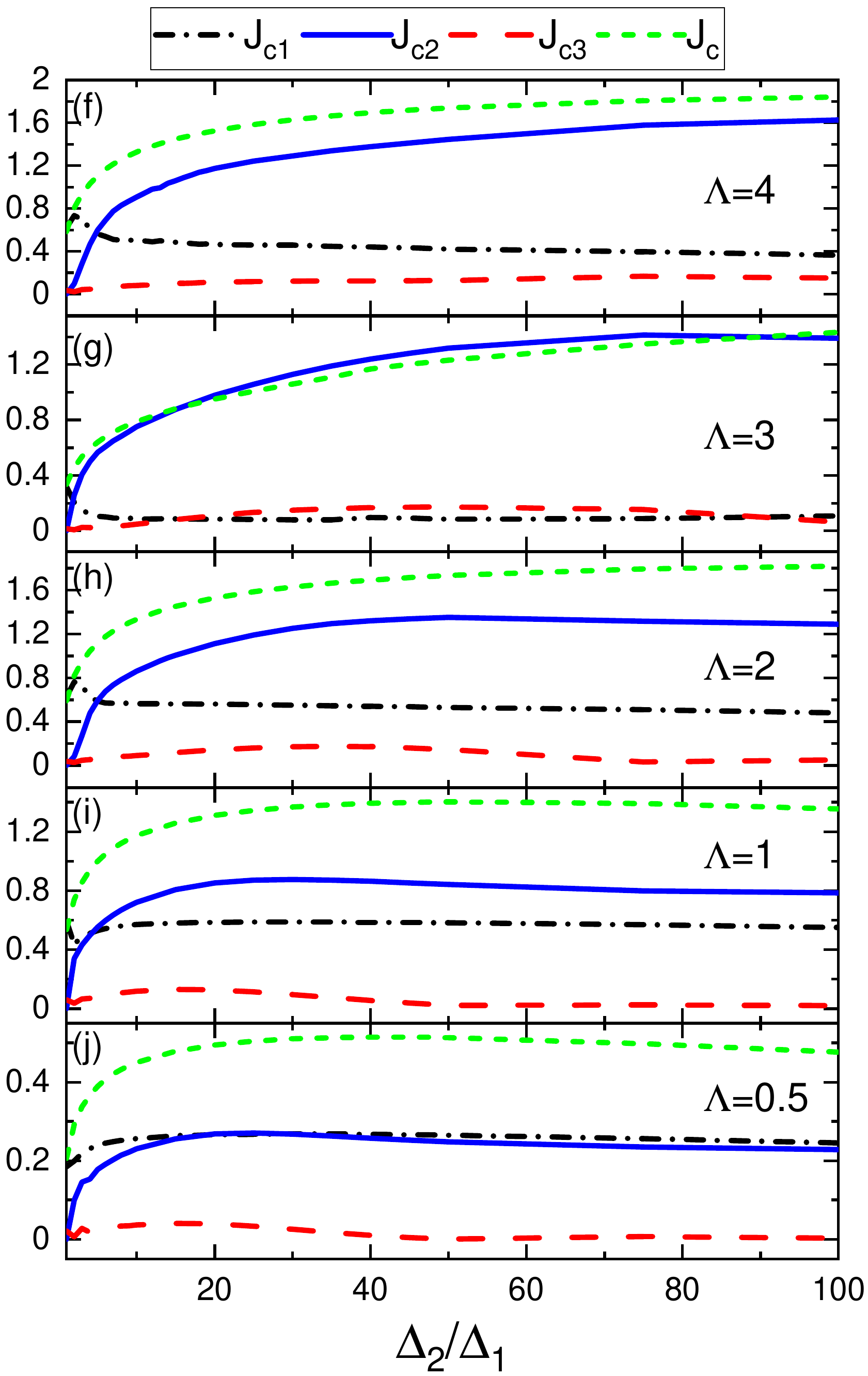}  }
\caption{(Color online). Critical supercurrent $J_c$ 
with its subgap $J_{c1}$, and supergap $J_{c2}, J_{c3}$ components 
shown as functions of the 
superconducting gap ratio $\Delta_2/\Delta_1$. The junction thickness is set  at $k_Fd_F=5$ for various 
values of $\Lambda=E_\text{FM}/E_\text{F}=0.5, 1,2,3,4$. The magnetization is set to zero 
in (a)-(e)  while $h/\Delta_1=15$ in panels (f)-(j). }
\label{fig3}
\end{figure*}

In Fig.~\ref{fig3},
we illustrate 
how $\Lambda$ affects
the behavior of the subgap ($J_{c1}$) and 
 the
supergap ($J_{c2},J_{c3}$) critical current components
as functions of \dd. Figures~\ref{fig3}(a)-\ref{fig3}(e)
correspond to the nonmagnetic \sns case ($h=0$),
while the right set \ref{fig3}(f) and \ref{fig3}(g)
  exhibits the critical current behavior for \sfs 
junctions with $h/\Delta_1=15$. 
Considering first the nonmagnetic  case, 
the left set of panels shows that
for a given gap ratio \dd, 
changing \La  
modulates  the  critical current, consistent with the findings shown in Fig.~\ref{fig5}.
When \La$=3$, the subgap supercurrent is strongly  
suppressed and the supergap current dominates the net behavior of the
critical current 
throughout the wide range of \dd considered. 
Indeed, Andreev bound states with energies less than $\Delta_1$
play little, if any role in the establishment of a supercurrent.
For \La$=2,4$, the subgap components all
have the same trends, including 
the subgap supercurrent $J_{c1}$, which
exceeds the supergap currents for \dd$\lesssim 5$,
after which it decays considerably.
The supergap current $J_{c2}$ on the other hand, rapidly increases
in the region  \dd$\lesssim 25$, surpassing $J_{c1}$ at \dd$\approx 10$,
before eventually leveling out
close to the total critical current curve.
Thus,
for  \dd$\gtrsim 25$,  only energies
that fall within $\Delta_1 < \epsilon < \Delta_2$ are needed
when calculating 
the critical supercurrent response,  while the Andreev bound states
with $\epsilon<\Delta_1$ and scattering states with $\epsilon>\Delta_2$
can
 be neglected.
It is evident that there is an intricate 
and nontrivial relationship 
between the Fermi level mismatch and the critical current.
For situations where there is no mismatch $\Lambda=1$,
Fig.~\ref{fig3}(d) shows that  the influence of 
$J_{c1}$ in this regime is significant, and cannot be neglected for
most gap ratios.
In particular, 
for
\dd$\leq50$,
the  $J_{c1}$ component  
exceeds all other components 
before slowly decaying at
higher values of \dd.
In Fig.~\ref{fig3}(e),  where $\Lambda=0.5$,
the crossover point occurs at the much smaller
\dd$\approx 6$,
 indicating
 that both $J_{c1}$ and $J_{c2}$ must be accounted for,
 even for moderate \dd ratios.
 These results indicate that when 
 characterizing the supercurrent decomposition,
 the degree of Fermi level mismatch and gap mismatch 
play an important   role in which quasiparticle energies 
contribute to the supercurrent response. Also, a 
notable  
feature in Figs. \ref{fig3}(a) and \ref{fig3}(b) is that $J_{c2}$
sometimes  
exceeds the total supercurrent. 
This arises mainly due to  the subgap  $J_{1}$ and 
supergap $J_{2}$  currents 
flowing in opposite directions.
This important point shall be discussed further below.  

Next, when the central layer possesses 
a uniform magnetization, the adjacent panels in 
 Figs.~\ref{fig3}(f)-\ref{fig3}(j)
 reveal a clear modification to 
all three components of the critical current.
For each of the four cases of Fermi level mismatch,  
the component $J_{c1}$,
has an extremely slow decay for  gap ratios 
within
$\Delta_2/\Delta_1  \gtrsim 5$.
Thus within this regime, and
 for quasiparticle energies with $\epsilon >5 \Delta_1$,
 the critical current is insensitive to the relative 
 gap
 ratios 
 characterizing the superconducting leads. 
 When $\Delta_2/\Delta_1=1$,
 corresponding to the commonly used scenario
of no asymmetry in the gaps,
 the supergap component $J_{c2}$ vanishes.
 In Figs.~\ref{fig3}(f) and \ref{fig3}(h), when $\Lambda=4$ and $\Lambda =2$, respectively,
 the subgap component $J_{c1}$
 dominates all other critical current components for 
 relatively small gap ratios $\Delta_2/\Delta_1 \lesssim 2$.
 Both of
 the two components $J_{c1}$ and $J_{c2}$ contribute equally
 to the critical current for $\Delta_2/\Delta_1 \approx 5$,
 with $J_{c2}$ the largest contributor for larger gap asymmetry.
 The picture changes considerably 
 when  $\Lambda=3$,  as
  Fig.~\ref{fig3}(g)
  illustrates that the conventional Andreev bound states characterized by $J_{c1}$
  play a minor role in the net supercurrent behavior when $\Delta_2/\Delta_1 \gtrsim 2$.
  When the Fermi levels are the same in each segment of the junction
  [Fig.~\ref{fig3}(i)],
  the subgap component becomes more
  influential  with $J_{c1}$ and $J_{c2}$ crossing at $\Delta_2/\Delta_1 \approx 4.5$.
  Finally, in Fig.~\ref{fig3}(j), 
  we consider $E_{FM}<E_{F}$
  and  $\Lambda =0.5$.
  In this regime, 
  both the subgap and supergap components
  contribute nearly equally for most gap ratios $\Delta_2/\Delta_1 \gtrsim 15$,
  while less asymmetry again has $J_{c1}$ the larger
  of the components.  
  It should be noted that
the enhancement of 
each of the
current components against \dd occurs within a smaller interval compared to the \sns junction. 
   
\begin{figure*}[t!] 
\centering
{\includegraphics[width=8.8cm,height=14.cm]{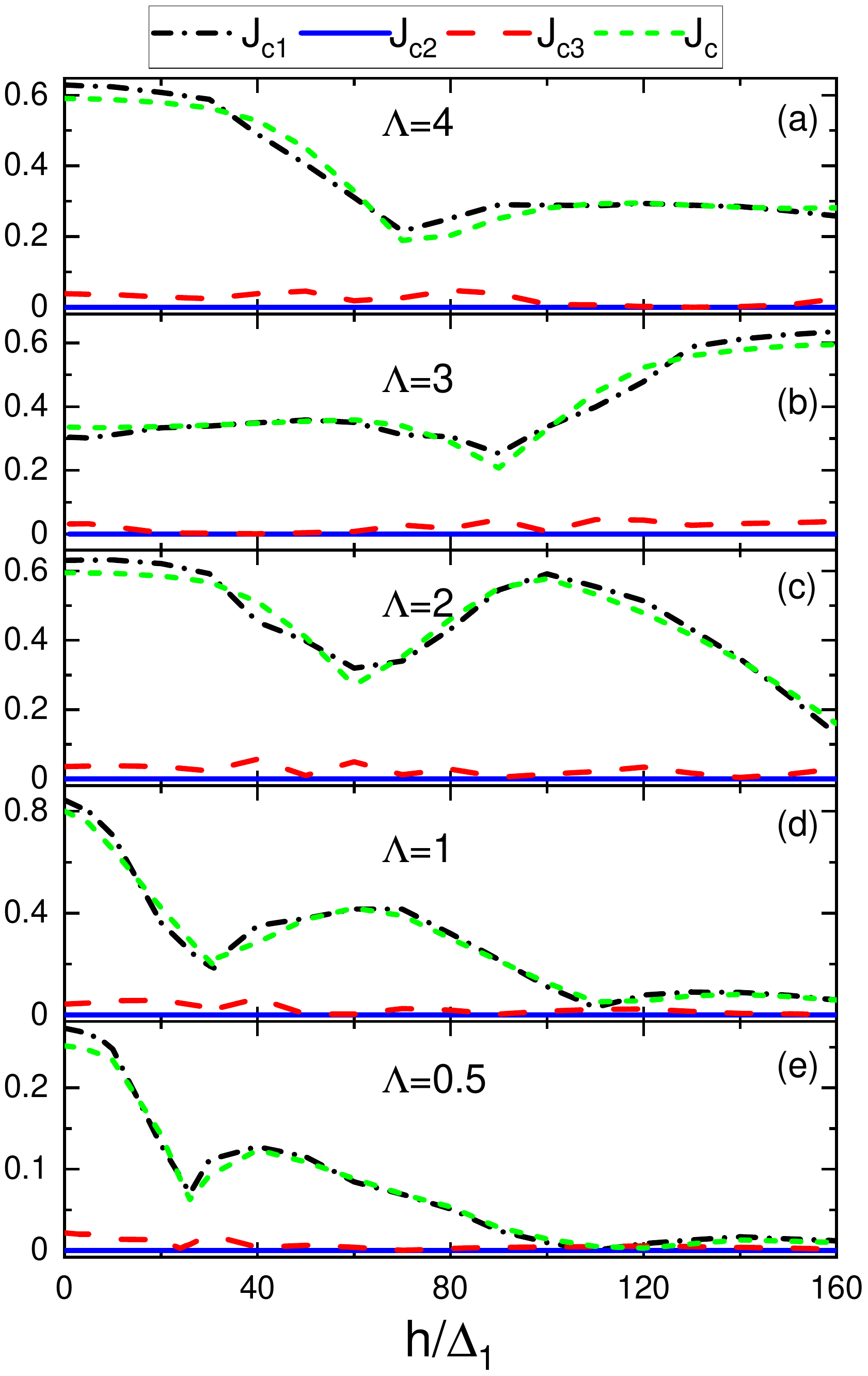} 
\includegraphics[width=8.8cm,height=14.cm]{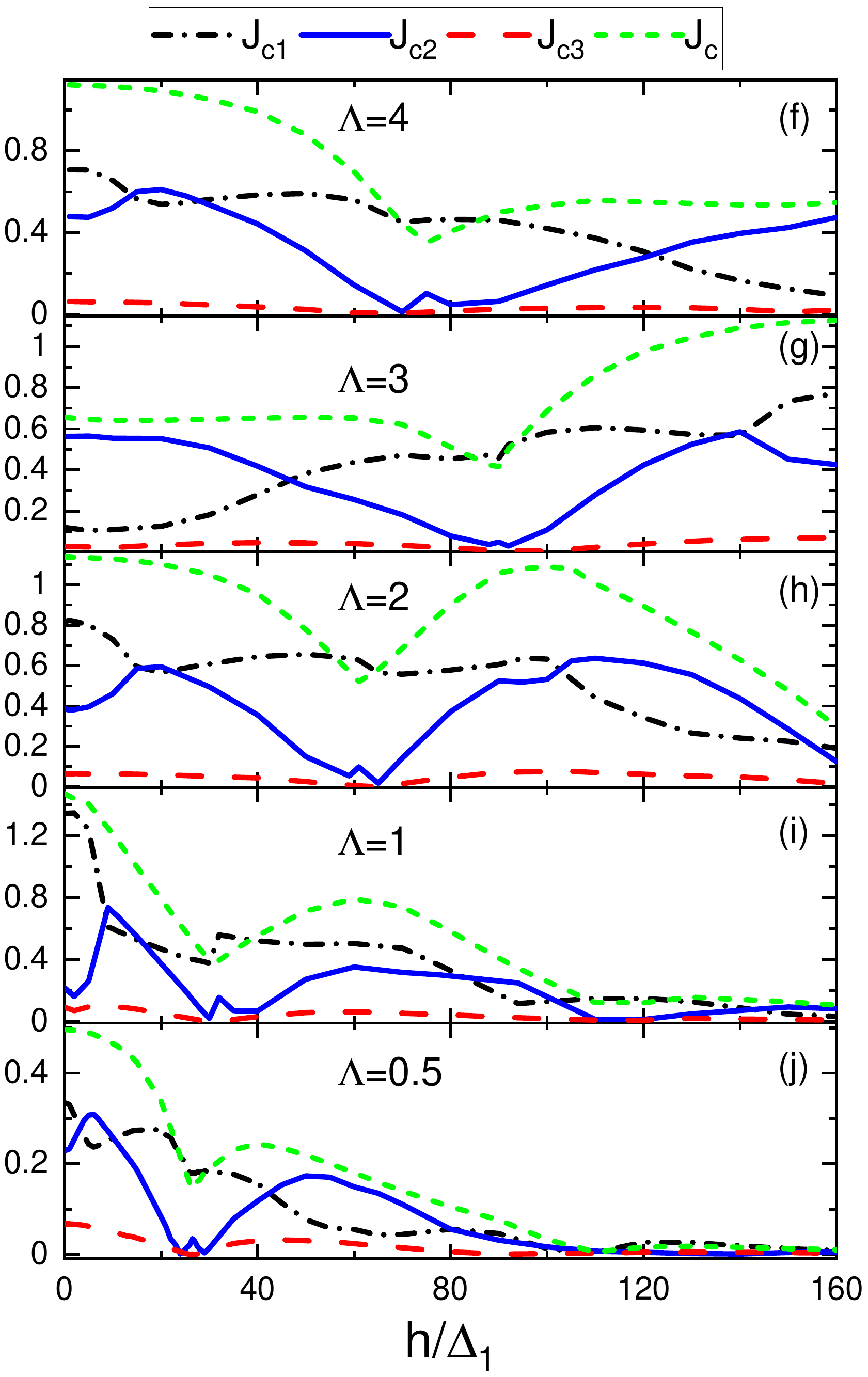}  }
\caption{(Color online). Critical supercurrent 
with its supergap and subgap components as a function of uniform magnetization strength $h/\Delta_1$. In the left column panels (a)-(e) the superconducting gap ratio is set to \dd$=1$ while in the right column panels (f)-(j) \dd$=5$.}
\label{fig4}
\end{figure*}

In conventional \sfs junctions,   
the exchange field induces damped oscillations in the
pair potential and
the Josephson current 
exhibits oscillations as 
a function of the ferromagnet exchange field.\cite{buzzed}
To see how this picture may change for
asymmetric junctions, we present in 
Fig.~\ref{fig4}  the behavior of the
critical supercurrent and its 
 components versus
the normalized exchange field $h/\Delta_1$. 
The study covers the full range
from nonmagnetic, $h/\Delta_1=0$, to
half-metallic, $h/\Delta_1 \approx 160$, where only one spin band is available.
Each panel corresponds to one of the five
different Fermi level mismatch ratios $\Lambda$ that are considered (as labeled).
To further clarify the importance of supergap and subgap supercurrents 
when there  is gap asymmetry,
we have considered 
 \dd$=1$ in Figs.~\ref{fig4}(a)-\ref{fig4}(e) and set \dd$=5$ in 
 Figs.~\ref{fig4}(f)-\ref{fig4}(j). 
 The results in Figs. \ref{fig4}(a)-\ref{fig4}(e) show that 
 for symmetric junctions (\dd$=1$), 
 the subgap supercurrent $J_{c1}$ is the dominant current 
 component over the entire range of magnetization strengths and  Fermi level mismatches. 
 The supergap contribution $J_{c2}$ to the total critical current of course vanishes entirely,
 while the   $J_{c3}$ component arising from states in the continuum is negligible.
 Hence, when
 the gap asymmetry vanishes (\dd$=1$), the behavior of 
 the total supercurrent can be well captured through $J_{c1}$ only.
It then suffices to
 take only
  the subgap current components
  to accurately account for the general features of the
  supercurrent, including its reversal in
certain regions where it displays cusps
  for a given $\Lambda$ and $h/\Delta_1$
  (see,  e.g.,  $h/\Delta_1\approx 30, 110$ in Fig. \ref{fig4}(d)). 
  Examining Figs.~\ref{fig4}(a)-\ref{fig4}(e), it is clear that the
  critical current is nonmonotonic with a modulation that strongly depends on $\Lambda$.
  Interestingly, tuning the Fermi level mismatch  to  $\Lambda=3$
  creates a situation where increasing the exchange field beyond the first cusp at $h/\Delta_1 \approx 90$
  results in a dramatic rise in the supercurrent response.
  Indeed, Fig.~\ref{fig4}(b) demonstrates that compared to a nonmagnetic junction ($h=0$),
  using a half-metallic insert ($h/\Delta_1\approx 160$) causes $J_c$ to nearly double.
  
 If   gap asymmetry is now introduced  into the system ($\Delta_2/\Delta_1~=5$),
  Figs. \ref{fig4}(f)-\ref{fig4}(j) show the
  emergence of the 
  $J_{c2}$  component, 
  which can at times
   make sizable  
   contributions  to  the  total supercurrent.
   It is evident  that the cusps in $J_c$ where
    the 
      current reverses, can be ascribed to
    the cumulative effects of  $J_{c1}$ 
    and $J_{c2}$,
    regardless of the Fermi level ratio \La. Therefore, when \dd$\neq 1$, the subgap current component alone
    is unable to provide an accurate and complete picture of the total supercurrent.
    Further details on the origins of
    both the subgap and supergap   supercurrent components
    in terms of their discrete energy spectra and energy-resolved
    supercurrents
    is given in Appendix~\ref{e-resolved_ballistic}.

\begin{figure}[t] 
\centering
\resizebox{0.49\textwidth}{!}{
\includegraphics{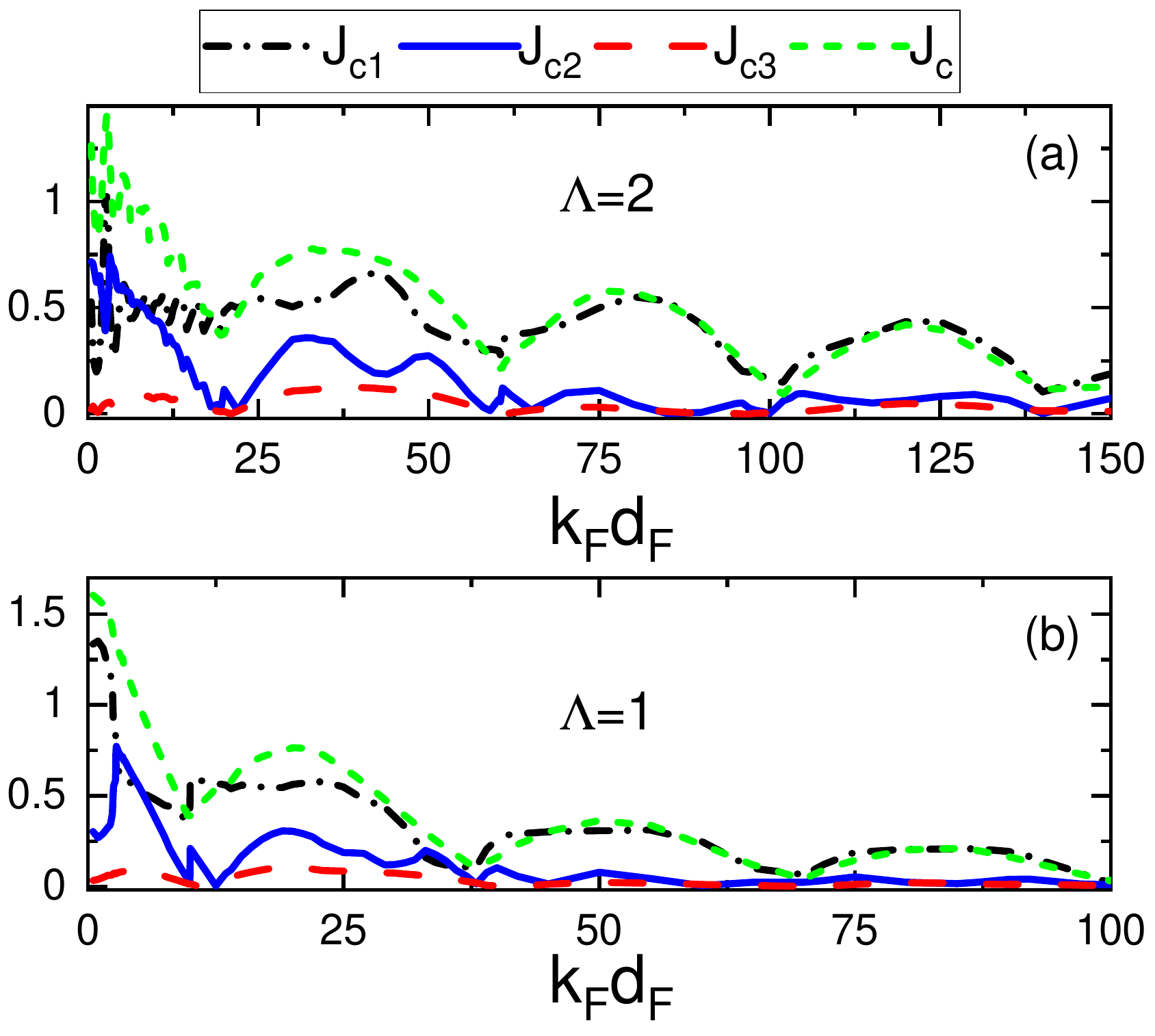}}
\caption{(Color online). Critical supercurrent with
its subgap and supergap components as a function of
the normalized junction thickness $k_Fd_F$ for an asymmetric  \sfs junction. 
The gap asymmetry for both panels (a) and (b)
corresponds to $\Delta_2/\Delta_1=5$ and the normalized 
exchange field is set to $h/\Delta_1=15$.
In (a) the Fermi level mismatch is set 
to $\Lambda=2$, while in (b) the relative Fermi levels are the same ($\Lambda=1$).
}
\label{figfull}
\end{figure}

 \begin{figure}[t] 
\centering
\resizebox{0.475\textwidth}{!}{
\includegraphics{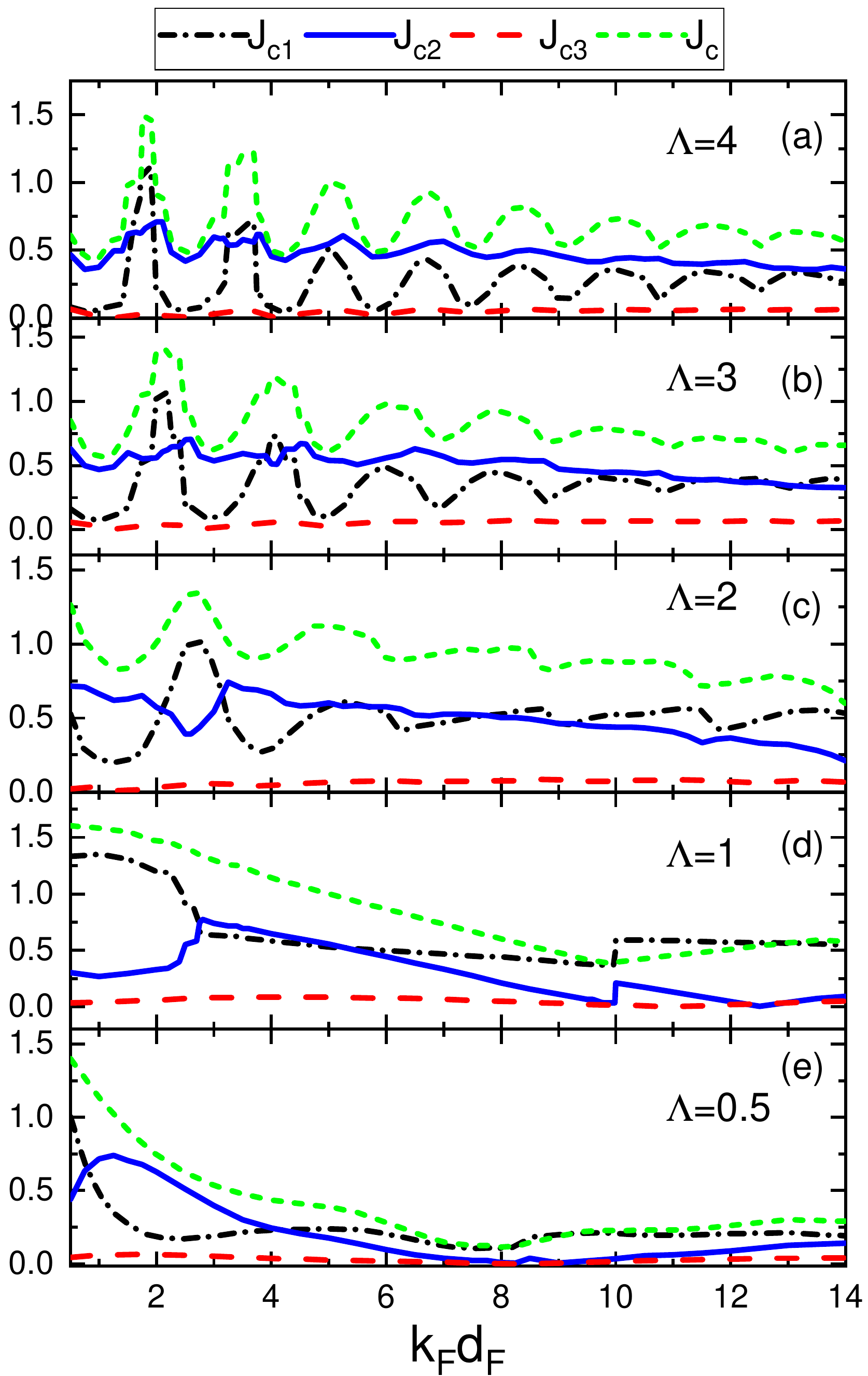}}
\caption{(Color online). Critical supercurrent with its subgap and supergap components as a function of junction thickness for differing chemical potential ratios \La$=0.5, 1, 2,3, 4$. The superconducting gap ratio and 
exchange field strength are set to \dd$=5$ and $h/\Delta_1=15$, respectively.}
\label{fig6}
\end{figure}    

\begin{figure*}[t!] 
\centering
{\includegraphics[width=0.49\textwidth]{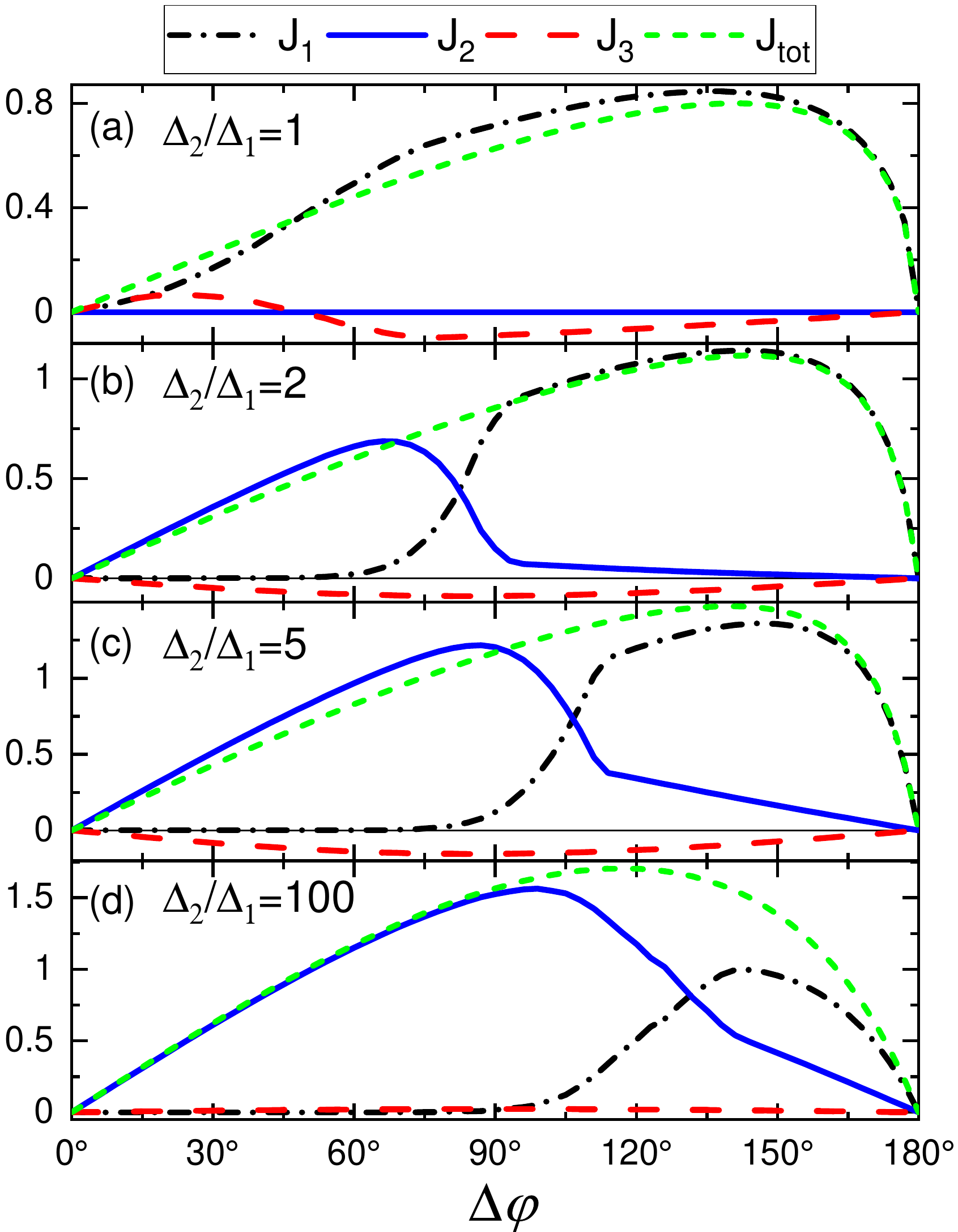} 
\includegraphics[width=0.49\textwidth]{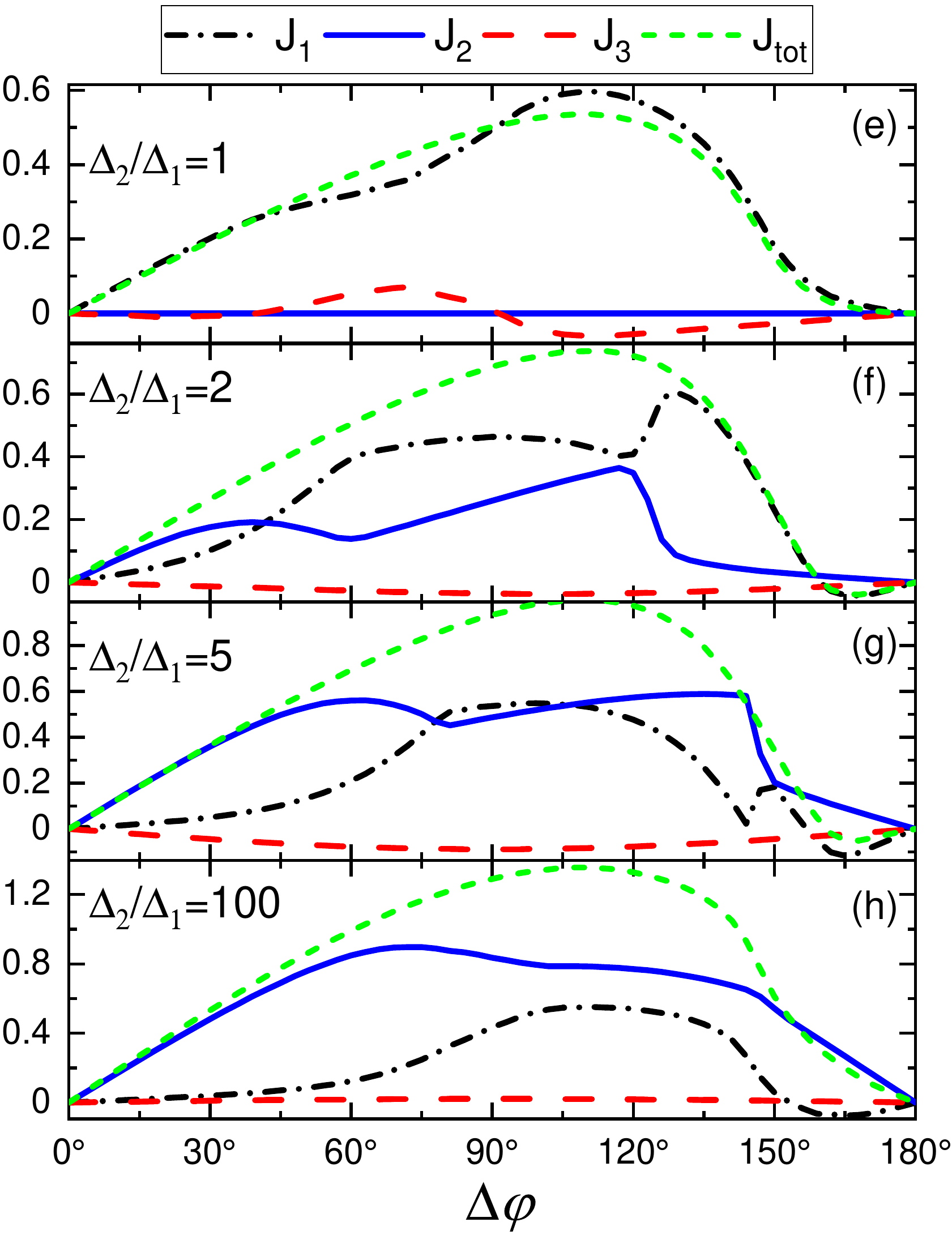}  }
\caption{(Color online).  Supergap ($J_{2,3}$) and subgap ($J_{1}$) current components as a function of 
the superconducting phase difference $\Delta\varphi$. (a)-(d): The superrcurent profile 
for a \sns junction. (e)-(h): The supercurrent profile for a \sfs junction with $h/\Delta_1=15$. 
The junction thickness is set fixed to $k_Fd_F=5$, and
the relative Fermi levels are equal ($\Lambda=1$).} 
\label{fig7}
\end{figure*}   

The  study of the critical current dependence on 
the ferromagnet thickness 
in \sfs Josephson junctions 
has been extensively investigated both theoretically\cite{buzzed} 
and experimentally\cite{kontos2002,shell2006}. 
        Proximity effects
        arising from the coupling of the 
        outer superconducting banks and
         ferromagnet
    leads to oscillations of the 
pair amplitude in the ferromagnet. 
Under certain conditions, these oscillations  can cause the
 ground state of the system
 to transition to
 a state with  $\Delta\varphi =\pi$.
 During these transitions,
 the supercurrent reverses direction and reveals itself 
 as cusps near the minima of the critical current.
Unfortunately, 
the study of transport in
clean ferromagnetic Josephson junctions
with Fermi level mismatch and gap asymmetry 
from a purely microscopic perspective is lacking. Therefore,
to address these deficiencies, we  present in
Fig.~\ref{figfull}, 
the critical current and its associated components as a function of normalized junction thickness $k_F d_F$.
The junction possesses a gap asymmetry of  $\Delta_2/\Delta_1=5$, and
two different values of the Fermi level mismatch parameter $\Lambda=1,2$ are considered.
Since the damped oscillations in the ferromagnet are governed by the
spin-split Fermi wavevectors there,
having $E_F$ vary across different segments of the Josephson junction
can modify the oscillatory period of the pair amplitude.
The spin splitting in the ferromagnet introduces the
length scale $\xi_F$ set by the difference in the spin up and spin down Fermi wavevectors,
$\xi_F \propto (k_{F\uparrow}-k_{F\downarrow})^{-1}$.
Accordingly, 
the cusps are found to repeat in intervals of $\pi\xi_F\approx\pi E_F \sqrt{\Lambda}/h$,
which for $h/\Delta_1=15$ and $\Lambda=1,2$, corresponds to  $\pi\xi_F \approx 32.9, 46.9$,
respectively.
This is seen when comparing Figs. \ref{figfull}(a) and \ref{figfull}(b).
Note that the damped oscillations in the critical current with thickness  have also
been observed in the critical temperature  for
ballistic spin valves.\cite{zep}

As seen in Fig.~\ref{figfull},
for a given $\Lambda$,
the supergap component $J_{c2}$  contributes
the most for small ferromagnet thicknesses.
To explore this further,
Fig.~\ref{fig6} displays the
critical current and its components 
over a narrower range of thicknesses
for the \sfs configuration.
To reveal how the oscillations and magnitude of
the supercurrent changes with
variations in the  Fermi level ratios,
a broad range of mismatch parameters $\Lambda$ is considered.
As seen in Figs. \ref{fig6}(a)-\ref{fig6}(e), the $J_{c3}$ component is negligible,
  as scattering states again contribute little
to the supercurrent response.
The 
total supercurrent oscillates as a function of 
$k_Fd_F$, but in contrast to Fig.~\ref{figfull},
these oscillations are over the much smaller Fermi length scale.
These  small-scale oscillations are neglected in quasiclassical treatments where 
atomic scale features are eliminated.
Moreover, increasing $\Lambda$
is shown to decrease the period of oscillations in Figs.~\ref{fig5}(a)- \ref{fig5}(e),
which is consistent with the corresponding increases in 
the ratios of the Fermi wavevectors in the  ferromagnet and superconductor regions $k_{FM}/k_F$.
For Fermi level ratios corresponding to $\Lambda = 4$ and $\Lambda=3$, 
Figs.~\ref{fig5}(a) and \ref{fig5}(b) illustrate that while $J_{c2}$
has gradual variations as the thickness changes,
the subgap component $J_{c1}$ has pronounced oscillations that periodically vanish (or nearly so)
for thin ferromagnets.
Therefore certain thicknesses can be chosen that result in maximal 
supercurrent flow that is almost entirely 
comprised of supergap $J_{c2}$ states.
As the difference between Fermi levels  lessens, Figs. \ref{fig5}(c) and \ref{fig5}(d)
show that the oscillatory behavior dampens out for thicker ferromagnets.
Finally in Fig. \ref{fig5}(e), we find that the critical current mainly declines rapidly
as
the normalized thickness increases and then  levels off to
greatly diminished values.

We now  present  the current-phase relations
 for the total supercurrent and its components
at specific parameter values  corresponding to  points of interest found in earlier critical supercurrent plots. 
 In Figs.~\ref{fig7}(a)-\ref{fig7}(d), we have set the 
 exchange field $h$ to zero and 
 consider  a \sns configuration with varying degrees of gap asymmetry.
  As seen in Fig.~\ref{fig7}(a), when there is no gap asymmetry ($\Delta_2/\Delta_1=1$),
 the $J_{3}$ component is slightly 
 discernible,
 as  states in the continuum contribute little to the supercurrent
  in short junctions.
 Also, the $J_{2}$ supergap component vanishes, as expected for the symmetric case.
 Thus the current in this case is carried nearly entirely by bound states.
 Examining Figs.~\ref{fig7}(b)-\ref{fig7}(d),
it is evident  that by increasing the asymmetry through the 
superconducting gap ratio, \dd, a supergap current $J_2$ 
 emerges, and begins to play a more impactful role in the
 total supercurrent response.
In contrast, 
the subgap current $J_1$ accordingly becomes 
weaker, as its contribution to 
at least 50\% of the total supercurrent is limited to relatively 
narrow  phase differences,
  $130^\circ\lesssim\Delta\varphi\lesssim 160^\circ$. 
Despite the significant changes to $J_1$ and $J_2$,
the overall supercurrent, $J_{\rm tot}$, increases, but retains its overall profile.
Increasing  $\Delta_2/\Delta_1$ also is seen to enhance the supercurrent overall, while
 shifting the supercurrent peaks to smaller $\Delta \varphi$.
 The enhancement of the supercurrent follows in part  from the broadening of the 
 discrete energy states that occurs for larger ratios $\Delta_2/\Delta_1$ [see Appendix~\ref{e-resolved_ballistic}].
The microscopic
 numerical results presented  in Figs.~\ref{fig7}(a)-\ref{fig7}(d) are also
 consistent with one-dimensional quasiclassical models\cite{bagwell}. 
 In Appendix~\ref{e-resolved_ballistic},
  Fig.~\ref{fig2D_sns}
  reveals  the interplay between
  the  bound and scattering states
related to Fig.~\ref{fig7},
and are discussed in terms of the 
energy-resolved and phase-resolved current density.

\begin{figure}[t] 
\centering
\resizebox{0.49\textwidth}{!}{
\includegraphics{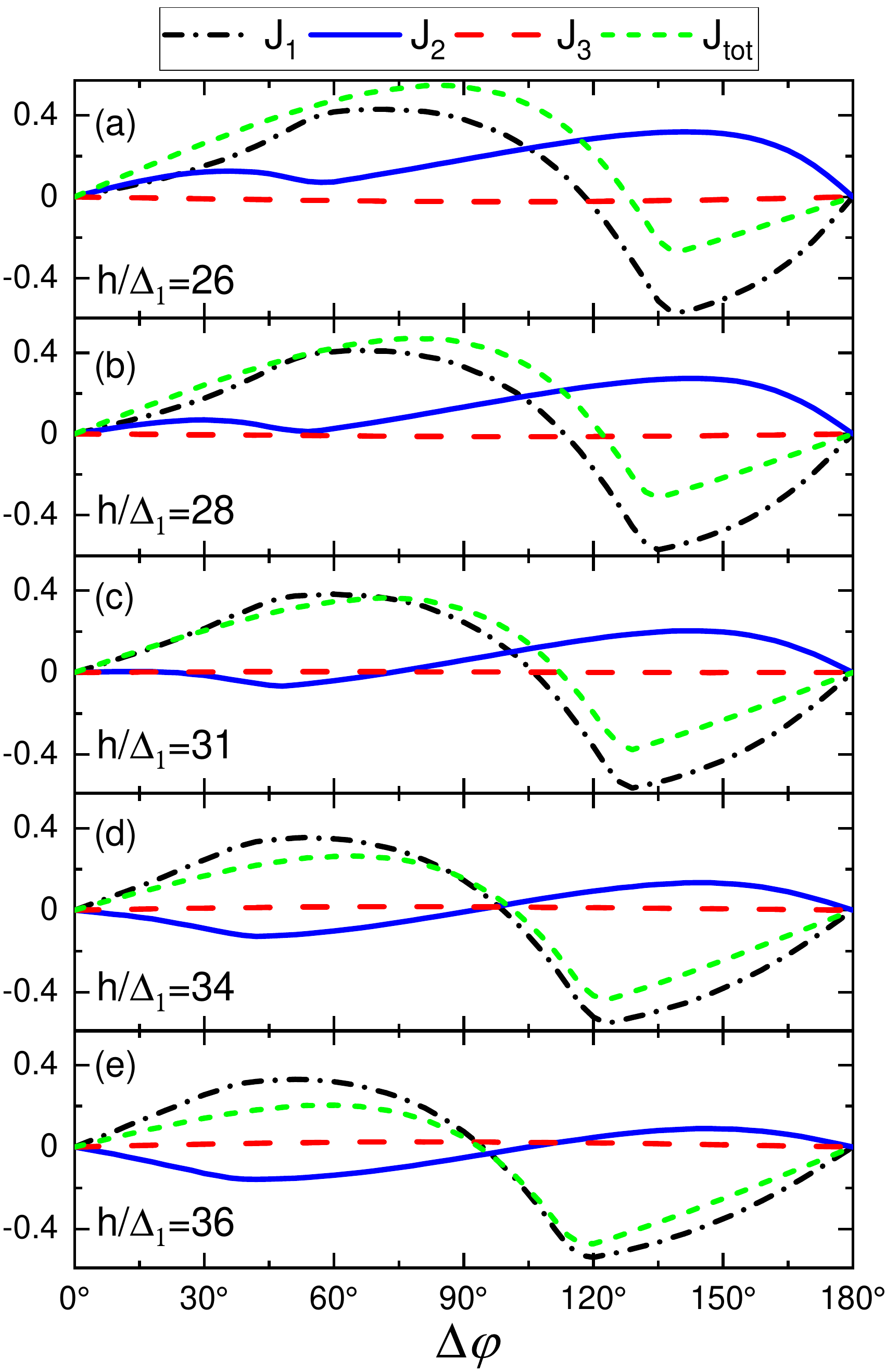}}
\caption{(Color online). Total current-phase relation with
 its supergap and subgap components for 
 several key exchange-field values around a current reversal point. The superconducting gap ratio is set to \dd$=5$,
  the junction thickness is $k_Fd_F=5$, and
 there is no mismatch in Fermi levels (\La$=1$).}
\label{fig8}
\end{figure} 

 Next, upon incorporating a uniform magnetization of $h/\Delta_1=15$, 
 Figs.~\ref{fig7}(e)-\ref{fig7}(h) illustrate how this leads to a
 drastic modification to the profiles of the current-phase relations.
 For 
  \sfs junctions, we find a  slight drop to the
  overall net supercurrent response 
  for the whole range of gap asymmetries considered.
  Increasing the gap asymmetry tends to enhance the overall supercurrent, but the general 
  current-phase profile remains relatively unchanged, with minimal
   change in the peak locations. One
   noticeable difference however occurs for $\Delta_2/\Delta_1=2,5$,
   where the supercurrent undergoes a current reversal at $\varphi \approx 162^\circ$.
  We find that similar to the symmetric \sns junction, Fig.~\ref{fig7}(e) shows 
  a small  contribution  from 
 $J_{3}$  arising  from scattering states ($\epsilon>\Delta_1$) 
 when there is gap symmetry.
  For extreme gap asymmetry $\Delta_2/\Delta_1=100$, shown in
  in Fig.~\ref{fig7}(h),  the supergap component $J_{2}$ is the 
  main contributor to the overall current, and we see that 
  although $J_{1}$ has broadened compared to the nonmagnetic case,
  it never exceeds approximately 40\% of the total critical current.
  
\begin{figure}[b]
\centering
\includegraphics[width=0.475\textwidth]{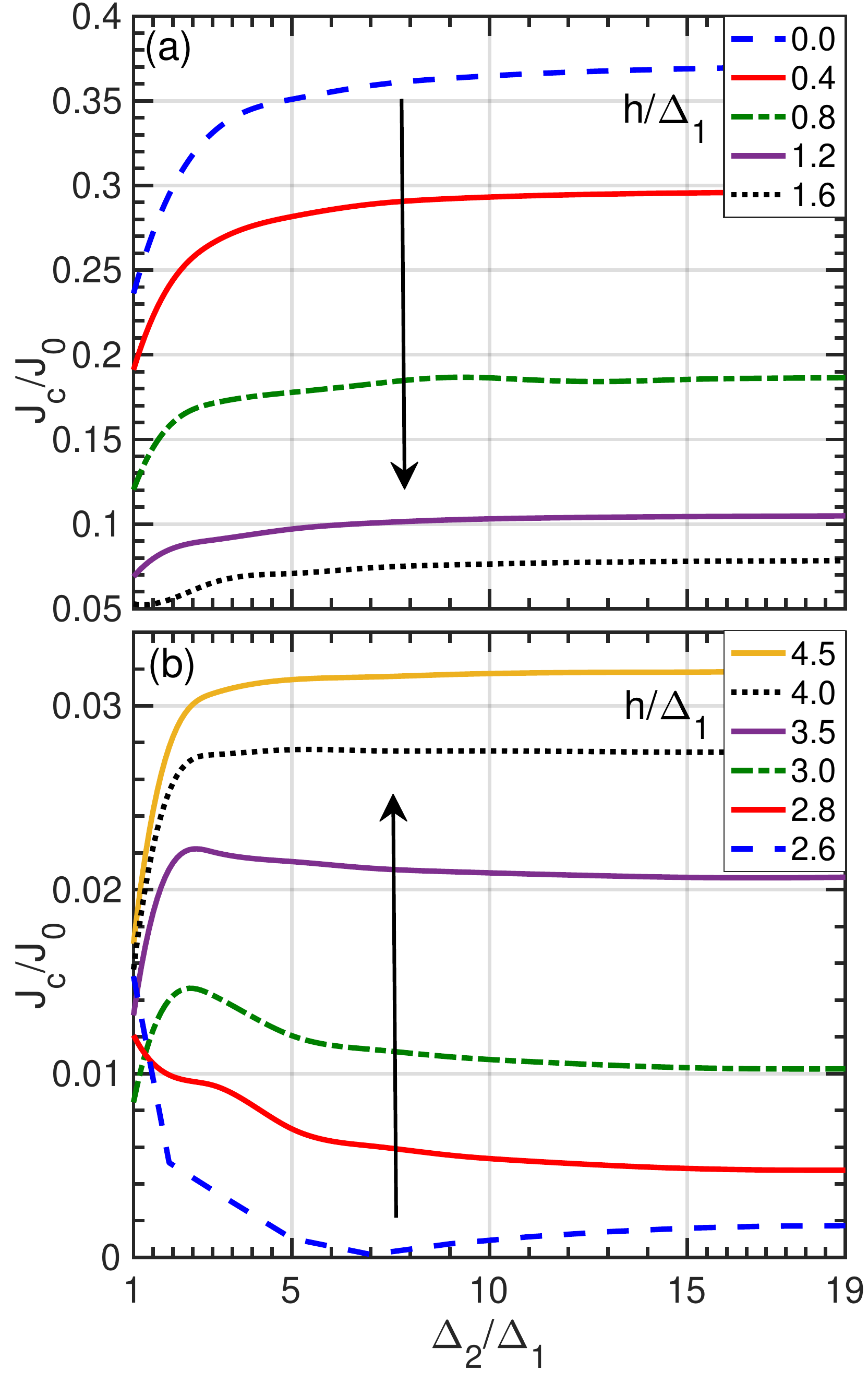}
\caption{(Color online). Critical supercurrent in a diffusive \sfs Josephson junction as a function of $\Delta_2/\Delta_1$. The junction thickness is fixed at $d=0.8\xi_S$ and various values of 
the normalized exchange field  are considered: 
$h/\Delta_1=0$, $0.4$, $0.8$, $1.2$, $1.6$, $2.6$, $2.8$, $3.0$, $3.5$, $4.0$, $4.5$.}
\label{Ic_diff}
\end{figure}   

\begin{figure*}
\centering
\includegraphics[width=1\textwidth]{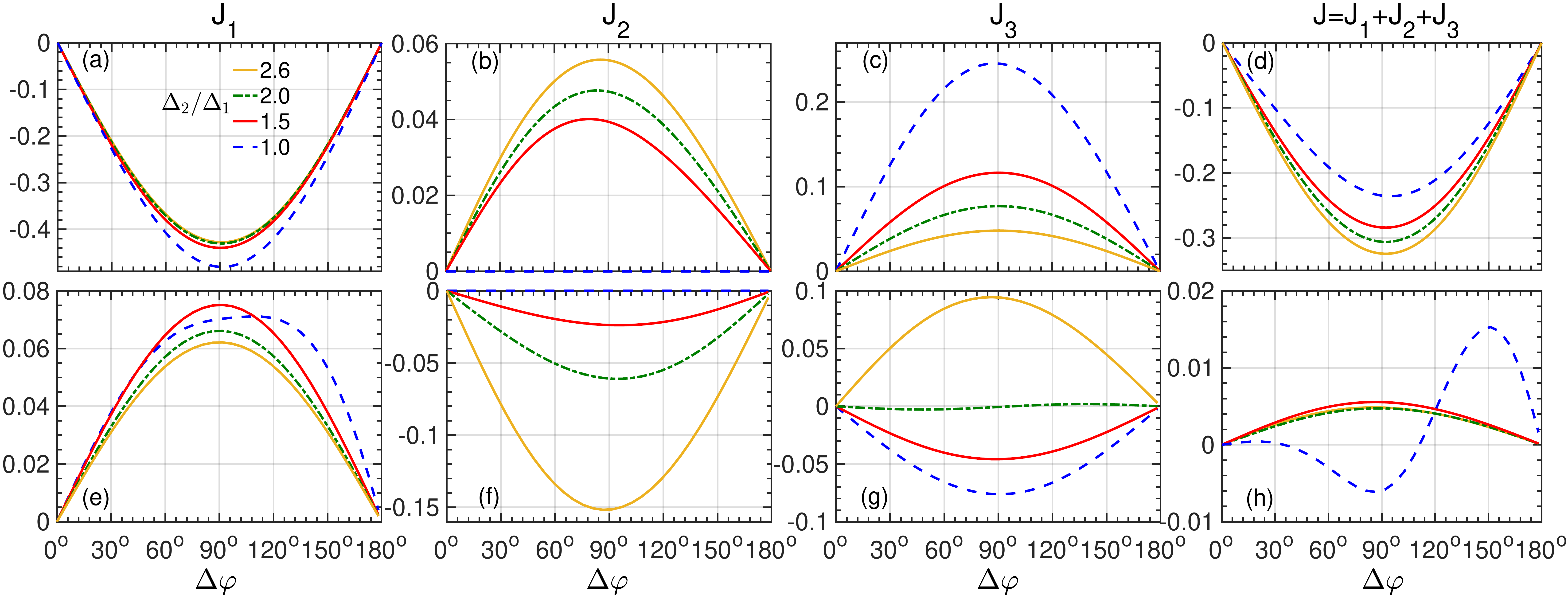} 
\caption{(Color online). The three components of supercurrent at four different values of 
the gap anisotropy ratio: $\Delta_2/\Delta_1$=$1.0$, $1.5$, $2.0$, and $2.6$. In (a)-(d), the exchange field $h$
 is set to zero, and  in (e)-(h), we set to $h=2.6\Delta_1$.}
\label{I123_diff}
\end{figure*} 
To elaborate 
on the
damped oscillations in the 
pairing correlations that induce
$0-\pi$ transitions 
responsible for 
  supercurrent reversals and cusps
in the critical current profile of
asymmetric junctions, we have studied the exchange field dependence 
of
the current-phase relations 
in Fig.~\ref{fig8}.
The total supercurrent $J_{\rm tot}$ is shown alongside its 
constitutive  components $J_{1,2,3}$. As a representative 
parameter set, we have chosen the first crossover state
in Fig.~\ref{fig4}(i), which 
occurs around $h/\Delta_1\approx 32$. Therefore, in 
Figs.~\ref{fig8}(a)-\ref{fig8}(e) we set \dd$=5$, \La$=1$, and $k_Fd_F=5$ and 
vary the normalized magnetization 
according to
$h/\Delta_1=26, 28, 31, 34, 36$, respectively. 
As seen, the overall profile 
and magnitude of the 
subgap supercurrent component $J_1$ is relatively unaffected 
as 
 the exchange field increases.
 The 
 supergap current $J_2$  however
 experiences a transition in which it goes from 
entirely  positive for $h/\Delta_1=26$ to negative
for most phase differences  $\Delta \varphi$
when $h/\Delta_1=36$.
This results in an overall
 suppression of the  total supercurrent, illustrating the influential control of
 the $J_2$ component on the  total supercurrent. 
 Therefore, adjacent to a $0-\pi$ crossover, the 
 supergap $J_2$ and subgap $J_1$ current 
 components propagate in opposite directions,
 creating a competing situation. 
 Below, we shall see that a similar occurrence 
 takes place near the
  $0-\pi$ transition point in the diffusive regime. 
 It should be noted that the supercurrent in the
 ballistic regime involves the superposition of quasiparticle trajectories  
 undergoing normal and Andreev reflections at the two superconductor interfaces. 
 Due to the  microscopic method used, length scales as small as the Fermi wavelength are included,
 permitting the capture of Friedel-like oscillations,\cite{sipr}
 which emerge as highly oscillatory signatures in the supercurrent response for small
 normalized  layer thicknesses $k_F d_F$. 
 The resonant modes are also highly sensitive to the relevant geometrical and material 
 parameters such as the junction length, Fermi level, magnetization strength, and superconducting gap ratio,
 all of which  intricately combine to give 
  the observed small oscillations in e.g., Figs.~\ref{fig6} and \ref{fig7}. 
  Nonetheless, the final conclusions and central findings are clearly independent of these subfeatures.

\subsection{Diffusive \sns and \sfs Josephson junctions}\label{results-diff}
We now consider the supercurrent response in asymmetric diffusive  Josephson junctions.
To properly capture   quasiparticles with energies  deemed relevant  to the net supercurrent, 
we have set an energy cutoff  of $\varepsilon_{max}=25\Delta_1$ 
when performing the integration over quasiparticle energies. 
We have also used 
 representative values for the junction thickness and interface opacity,
 with $d=0.8\xi_S$ and $\zeta=4$, respectively. 
 The maximum supercurrent flow 
  as a function of $\Delta_2/\Delta_1$ is shown  in  Fig.~\ref{Ic_diff}.
 The maximum of critical supercurrent occurs when the 
 magnetization in the junction vanishes. 
 As  seen, the critical supercurrent enhances
 by $50\%$ when $\Delta_2/\Delta_1\approx 10$ and $h=0$. Upon increasing the strength of the
  uniform exchange field $h$, the critical supercurrent 
  becomes suppressed.
 For weak ferromagnets with
   exchange energies  corresponding  to $h\sim 2.6\Delta_1$, the supercurrent undergoes a sign reversal as 
   the gap asymmetry $\Delta_2/\Delta_1$ is varied. 
   By further increasing $h/\Delta_1$ to $2.8, 3.0, 4.0, 4.5$, the overall critical supercurrent 
  is  enhanced and returns to its previous monotonic growth as a function of $\Delta_2/\Delta_1$. 
  In contrast to the other  normalized exchange field  strengths, the case
   with $h/\Delta_1=2.6$ exhibits a clear supercurrent suppression
   for $\Delta_2/\Delta_1\lesssim 7$.
   Increasing the exchange field to
 $h/\Delta_1=2.8, 3.0$, causes 
 the supercurrent to have a short-lived  enhancement,
 and a subsequent cusp at $\Delta_2/\Delta_1\approx 3$,
 before  slowly declining as the gap ratio increases. To create a  stable numerical scheme, 
 we have introduced a small imaginary part $\delta=0.001$ to the energy of the quasiparticles, $\varepsilon$. 
 This imaginary part can act as a source of inelastic scattering, which if increased, can wash out the dominant
 and  important parts of the curves, such as the overall supercurrent response 
 and associated $0$-$\pi$ transitions. For a relatively small imaginary part, 
 as is considered throughout our numerical study, the  algorithm 
 can at times introduce insignificant artifacts that  show up as small oscillations in the supercurrent.

To gain a comprehensive picture of the  
supercurrent in a diffusive asymmetric 
Josephson junction, we have plotted the
components  $J_{1,2,3}(\Delta\varphi)$, and total supercurrent $J(\Delta\varphi)$ in Fig.~\ref{I123_diff}. 
Various levels of gap asymmetry are shown
corresponding to  $\Delta_2/\Delta_1=1.0, 1.5, 2.0, 2.6$. 
The nonmagnetic \sns case is shown in
Figs.~\ref{I123_diff}(a)-\ref{I123_diff}(d),
 whereas Figs.~\ref{I123_diff}(e)-\ref{I123_diff}(h) 
 correspond to a \sfs junction. 
As seen for both cases, when $\Delta_2=\Delta_1$, the 
$J_2(\Delta\varphi)$  component vanishes, as expected. 
Also, there is an overall reduction  of the supercurrent magnitudes
in the magnetic case.
Other than their magnitudes, the 
current-phase relations of the 
 $J_2(\Delta\varphi)$ components are similar for
 \sns and \sfs 
  junctions with asymmetric superconducting gaps (albeit with different signs). 
  The subgap supercurrent $J_1(\Delta\varphi)$ behaves slightly differently for 
  \sns and \sfs junctions. In the latter case, when $\Delta_2=\Delta_1$,
   the supercurrent deviates strongly from the usual sinusoidal relation. In contrast to the
   \sns case, the component $J_3(\Delta\varphi)$ shows a
   $\sim\sin 2\Delta\varphi$ relation around $\Delta_2\approx 2\Delta_1$,
    and changes sign for larger values. The competition between
    these components results in the
    total supercurrents shown in the far right panels. As is apparent, unlike the strikingly
     different responses of $J_{1,2,3}(\Delta\varphi)$ for $\Delta_2> \Delta_1$, the total supercurrent 
      changes uniformly except when transitioning from  $\Delta_2= \Delta_1$ to $\Delta_2> \Delta_1$ for
      the
      \sfs
      junction with $h=2.6\Delta_1$. This variation results in the form of the
      current-phase relation changing  from $\sim\sin 2 \Delta\varphi$ to $\sim\sin \Delta\varphi$.
       Note that the total supercurrent response  in Fig.~\ref{I123_diff}(h) 
       has the form $\sim\sin 2 \Delta\varphi$, which
      appears due to the competition between the subgap supercurrent states comprising 
      $J_1(\Delta\varphi)$ and 
      the scattering states which embody $J_3(\Delta\varphi)\sim\sin \Delta\varphi$. The competition originates from the opposite propagation directions of the $J_1(\Delta\varphi)$ and $J_3(\Delta\varphi)$ current components. 
      These findings thus
      complement the ballistic results that
      found many instances where
       supergap modes must be accounted for
       appropriately to obtain accurate and reliable results.  
       Further insight into the supergap and subgap responses 
        are presented in  Appendix~\ref{e-resolved_diff},
        where the energy-resolved and phase-resolved 
        supercurrent density is analyzed.

\section{Conclusions}\label{conclusions}
By employing complementary
numerical approaches in the ballistic and diffusive regimes, we have performed a comprehensive study of 
 supercurrent flow through asymmetric \sns and \sfs Josephson junctions where the superconducting gap in 
 the $\rm S_{1,2}$ regions are unequal, i.e., $\Delta_2\neq \Delta_1$. In the ballistic regime, we have 
 directly solved the
 Bogoliubov de-Gennes Hamiltonian that allows for exploring
 a parameter space with a wide range of energy and length scales, whereas when  impurities and disorder
 are present, 
 we make use of the full proximity limit of the
  quasiclassical regime. Our results found that 
  for asymmetric junctions with  \dd$\approx 25,10$,
  the
  critical supercurrent can be enhanced by more than $100\%$ and $50\%$ in the ballistic and diffusive \sns cases, respectively. Our results in the ballistic
  cases reveal that when \dd$=1$, 
  the
  subgap current is the main contributor to total supercurrent.
   Introducing an imbalance to the superconducting gap ratio \dd$> 1$, the
  supergap currents were discovered to
   play key roles and for certain parameter values 
   were the main contributors  to
  the  total supercurrent. Through our investigations of  asymmetric junctions (with \dd$>1$), the
  current phase relations with their supergap and subgap current components were explored around  $0-\pi$ current crossover points.
   We  found that the emergence of second harmonics in the
   current-phase relations of  
   \sfs junctions is a direct consequence of the competition between subgap and supergap current components with opposite flow directions.    
   It  was shown in an earlier work\cite{bagwell} that 
 supergap currents are relatively insensitive to 
temperature compared to the subgap component,
 as the former originates from  coherent evanescent 
 modes in the continuum, whereas the latter is carried 
 through resonant bound states. 
 Therefore, the findings of this paper should serve to 
stimulate  experiments
that pave the way for
designing  new
superconducting devices  
that utilize 
  robust supergap currents.
The asymmetric \sns and \sfs structures
studied here can  apply to Josephson configurations 
where the amplitude of the superconducting gaps might fluctuate independently when the system is subject to a strong external magnetic field or high  temperatures near the  critical temperature. Furthermore, the enhancement of 
the critical supercurrent due to $\Delta_1\neq\Delta_2$ can be beneficial in magnetic Josephson junctions that suffer from weakened currents in the presence of ferromagnetism.

\acknowledgments
M.A. is supported by Iran’s National Elites Foundation (INEF). 
K.H. is supported in part by ONR and a grant of HPC resources from the DOD HPCMP.

\appendix

\begin{figure*}[t!] 
\centering
{
\includegraphics[width=0.244\textwidth,scale=0.01]{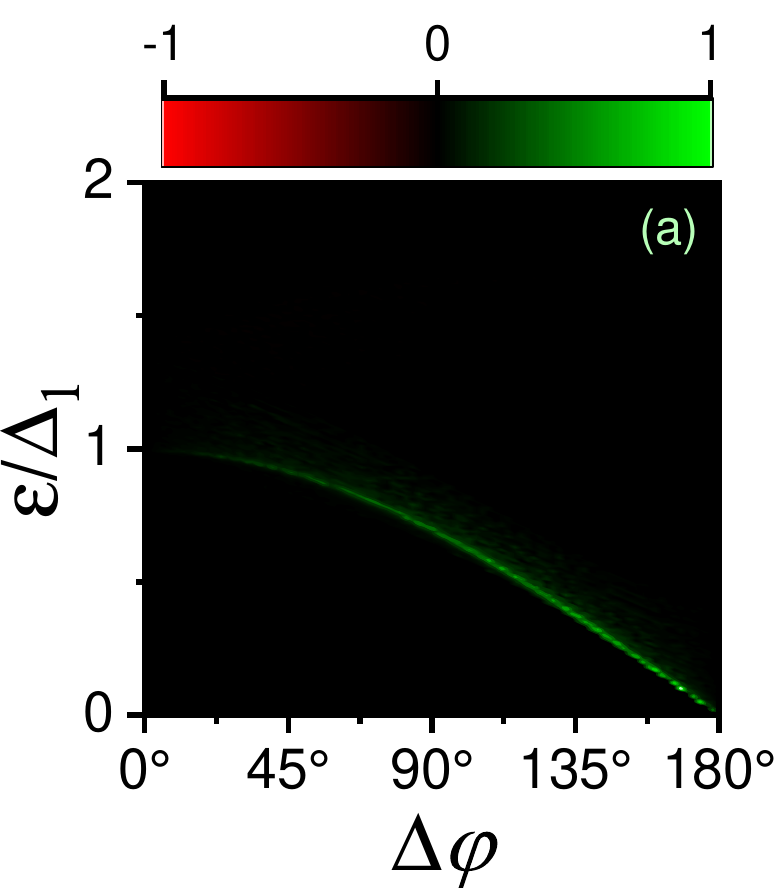} 
\includegraphics[width=0.244\textwidth,scale=0.01]{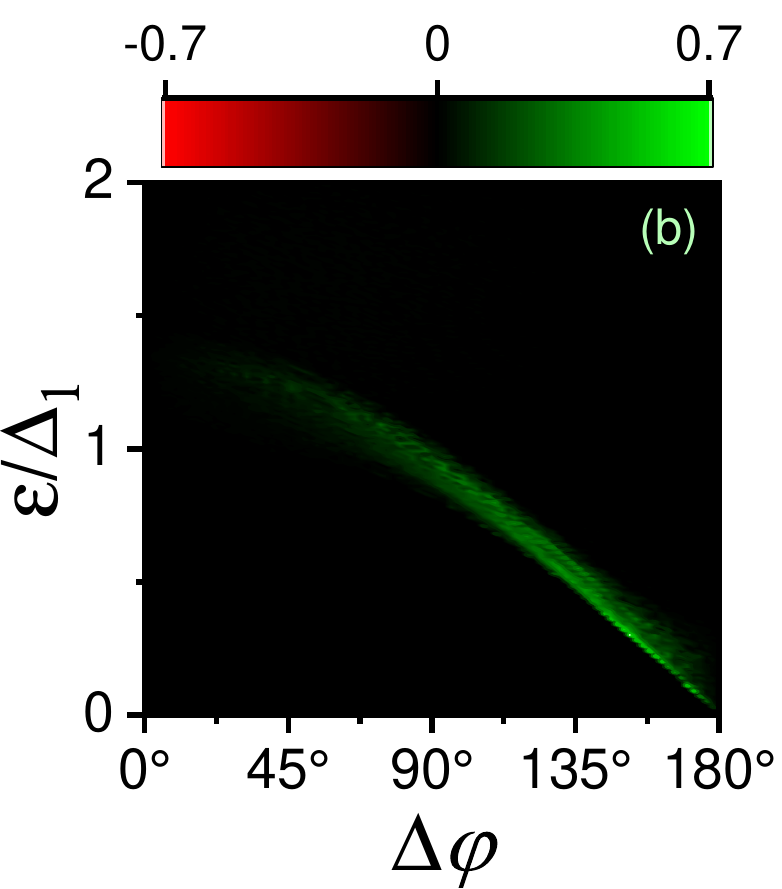} 
\includegraphics[width=0.244\textwidth,scale=0.01]{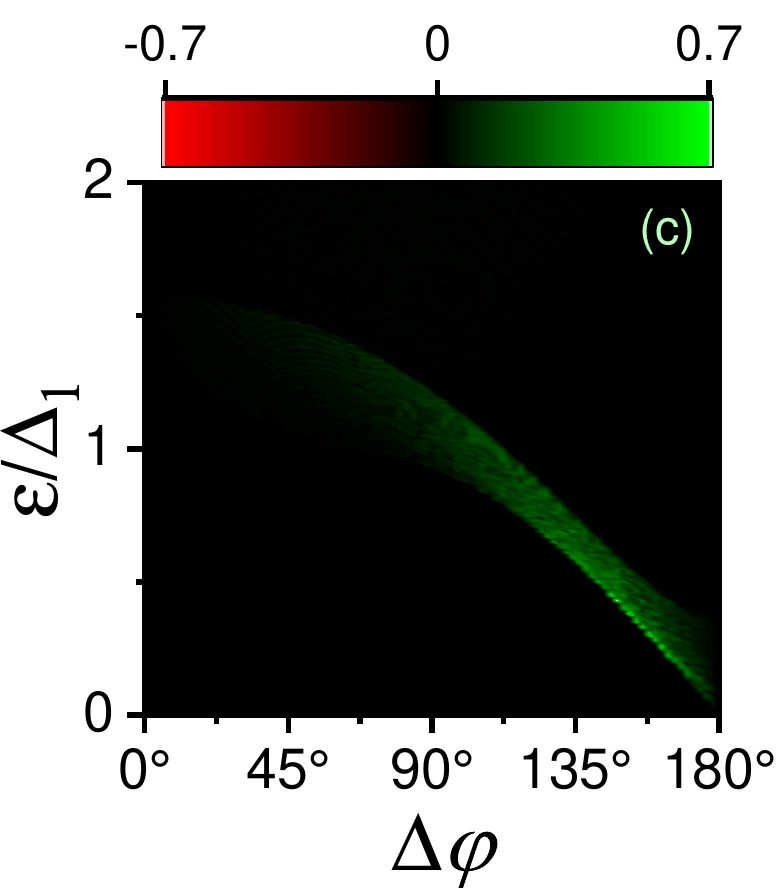} 
\includegraphics[width=0.244\textwidth,scale=0.01]{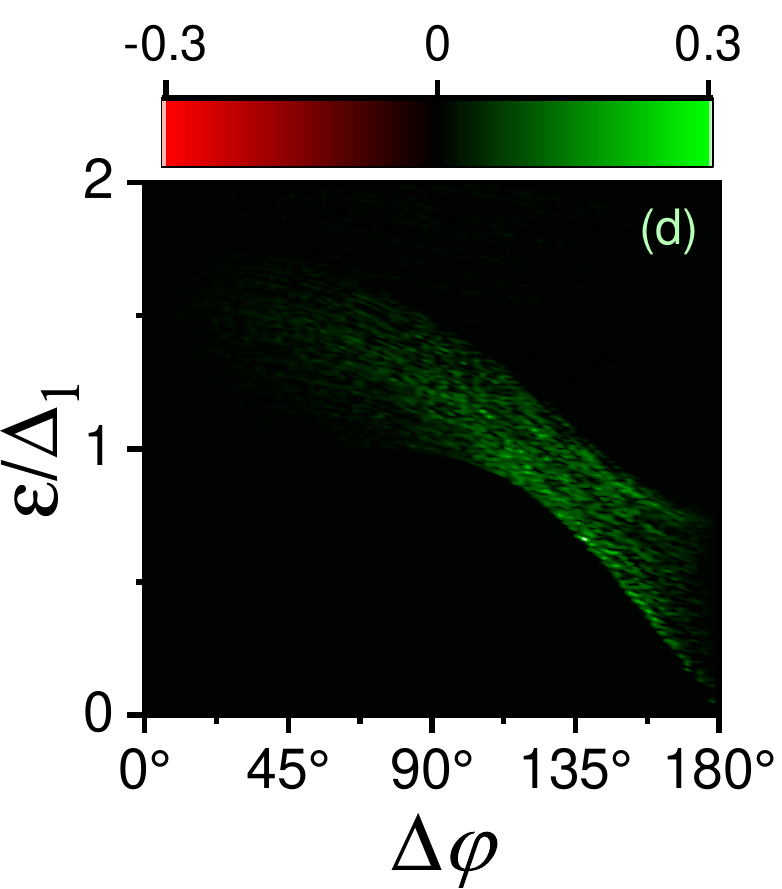} 
 }
\caption{(Color online). Total supercurrent density mappings as a function of phase difference $\Delta\varphi$ and 
energy $\varepsilon$ for a \sns Josephson junction with 
varying  levels of gap  asymmetry:
(a) $\Delta_2/\Delta_1=1$, (b) $\Delta_2/\Delta_1=2$, (c) $\Delta_2/\Delta_1=5$,
and  (d) $\Delta_2/\Delta_1=100$.
The Fermi levels are equal throughout the system ($\Lambda = 1$).
 }
\label{fig2D_sns}
\end{figure*}

\begin{figure}[b!] 
\centering
\resizebox{0.475\textwidth}{!}{
\includegraphics{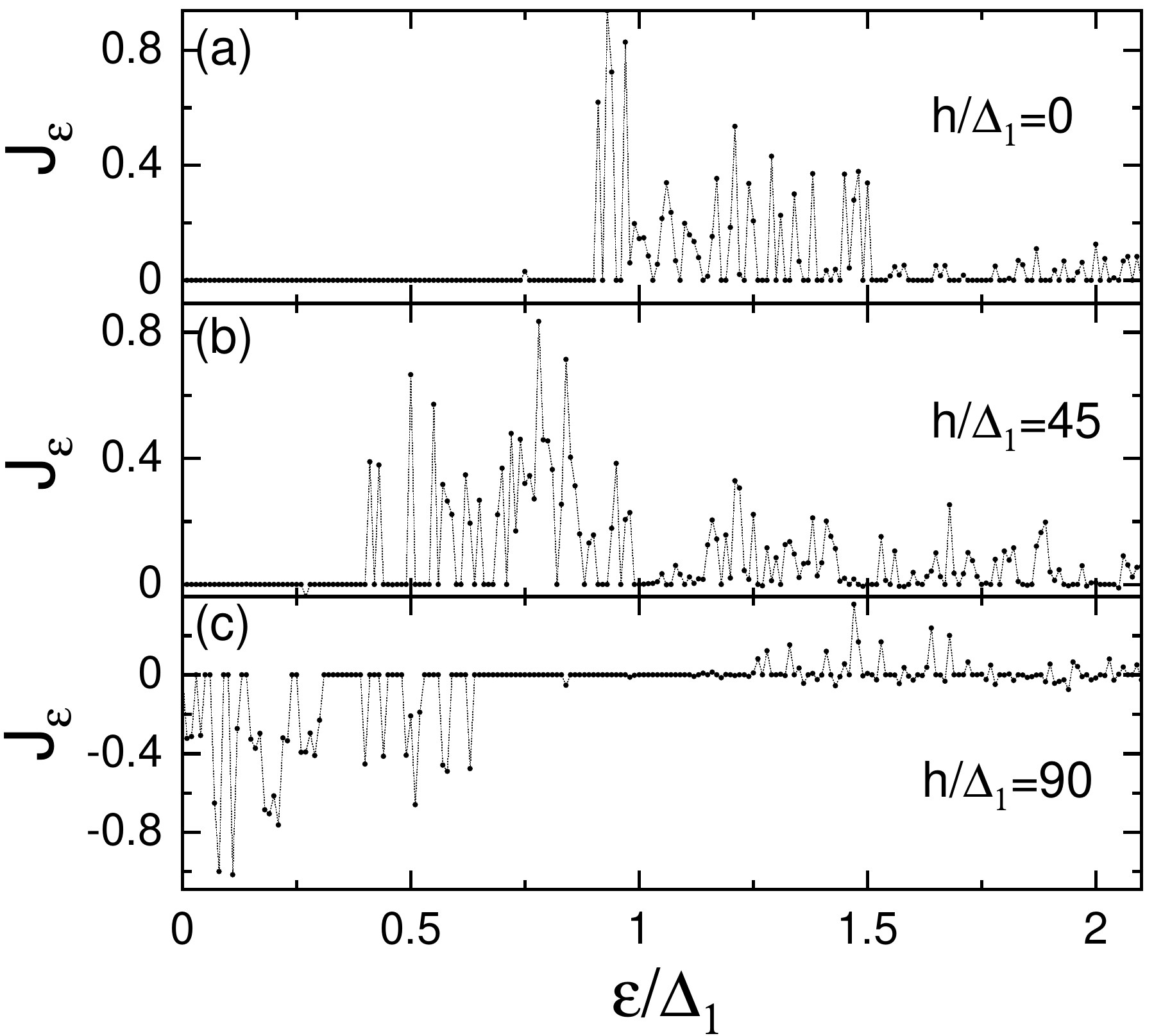}}
\caption{
(Color online). The energy-resolved supercurrent density $J_\varepsilon$ for 
an \sfs junction with $\Delta_2/\Delta_1=5$,
and Fermi level mismatch $\Lambda=3$. Three different values of the normalized exchange 
energy $h/\Delta_1$ are considered:
(a) $h/\Delta_1=0$, (b) $h/\Delta_1=45$, and (c) $h/\Delta_1=90$.
The macroscopic phase difference chosen in each case corresponds to $\Delta\varphi_c$,
where $\Delta\varphi_c$ is the phase angle that leads to the largest magnitude of
the total supercurrent. In panels (a) and (b)  $\Delta\varphi_c=106^\circ$,
and for (c) we have $\Delta\varphi_c=125^\circ$.
}
\label{jspike}
\end{figure}   

\begin{figure*}[t!] 
\centering
{
\includegraphics[width=0.24\textwidth,scale=0.01]{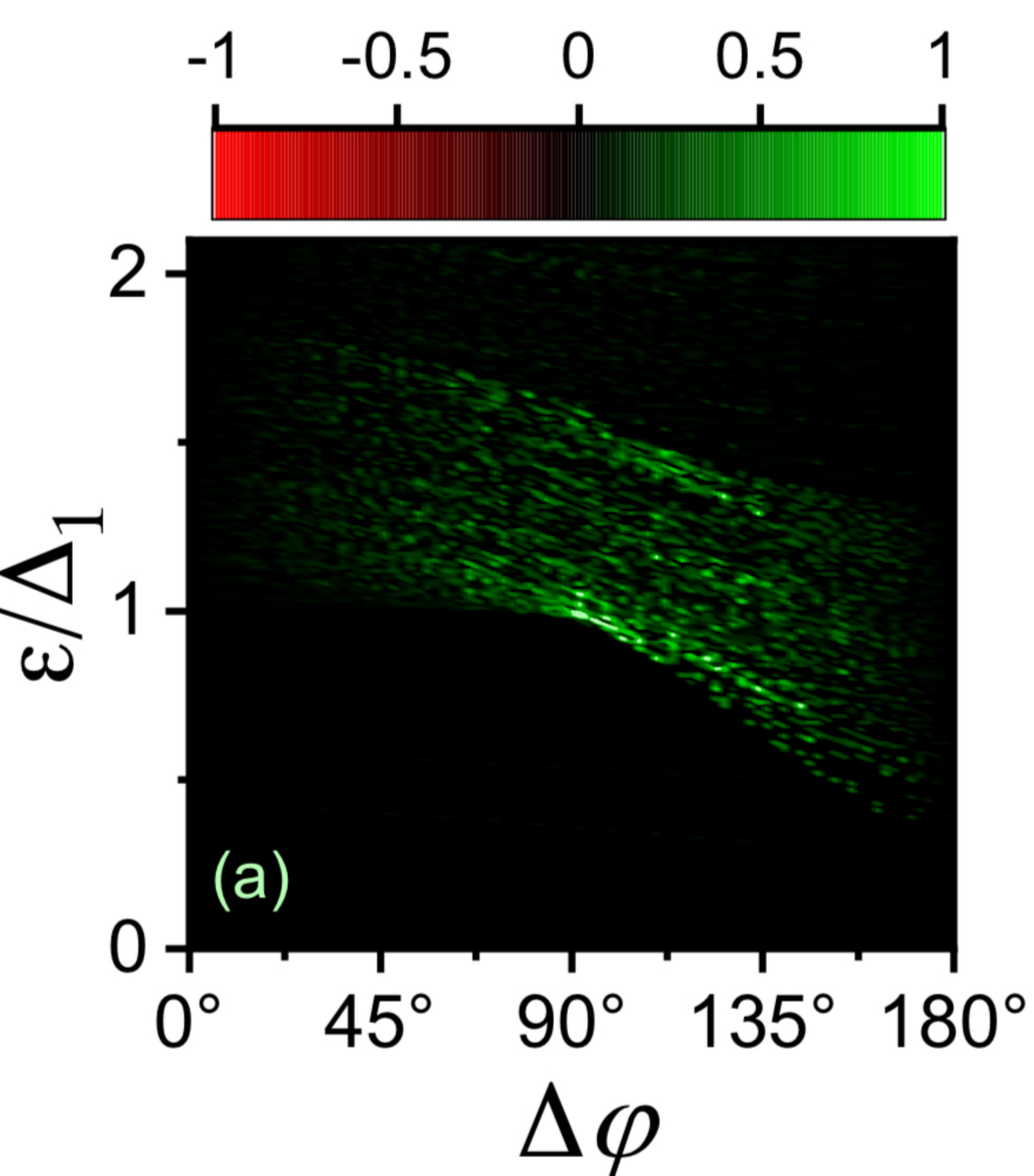} 
\includegraphics[width=0.24\textwidth,scale=0.01]{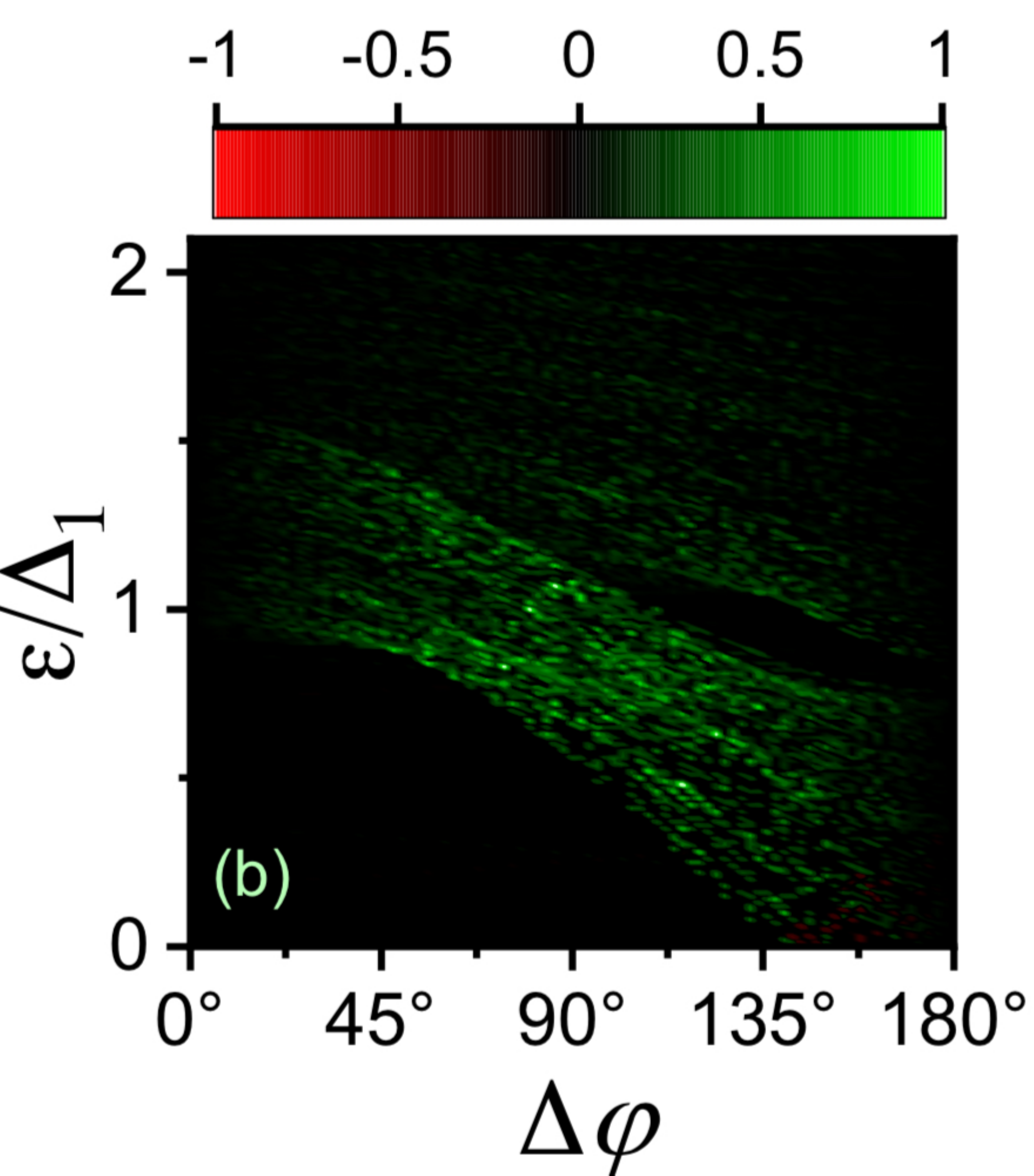} 
\includegraphics[width=0.24\textwidth,scale=0.01]{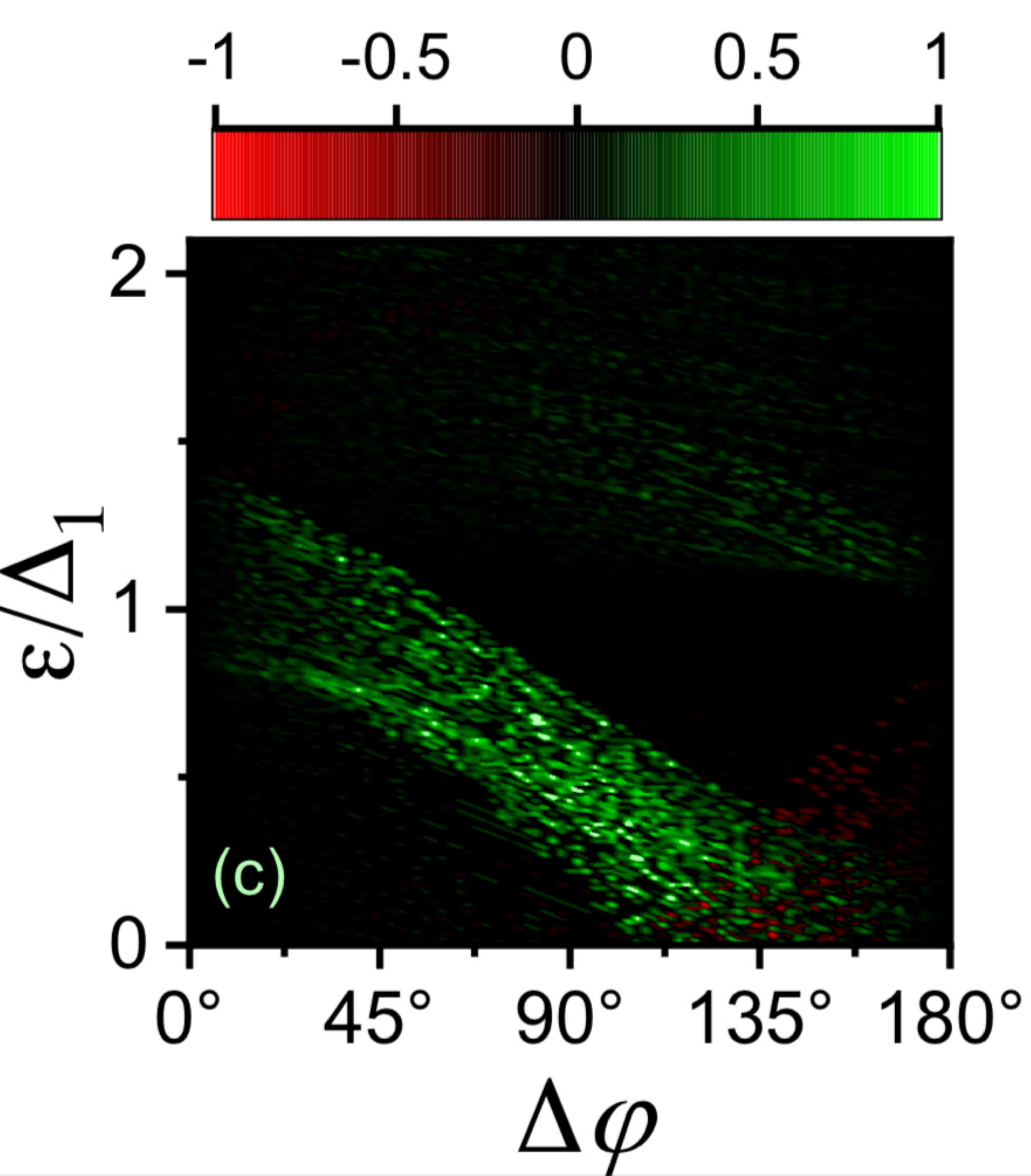} 
\includegraphics[width=0.24\textwidth,scale=0.01]{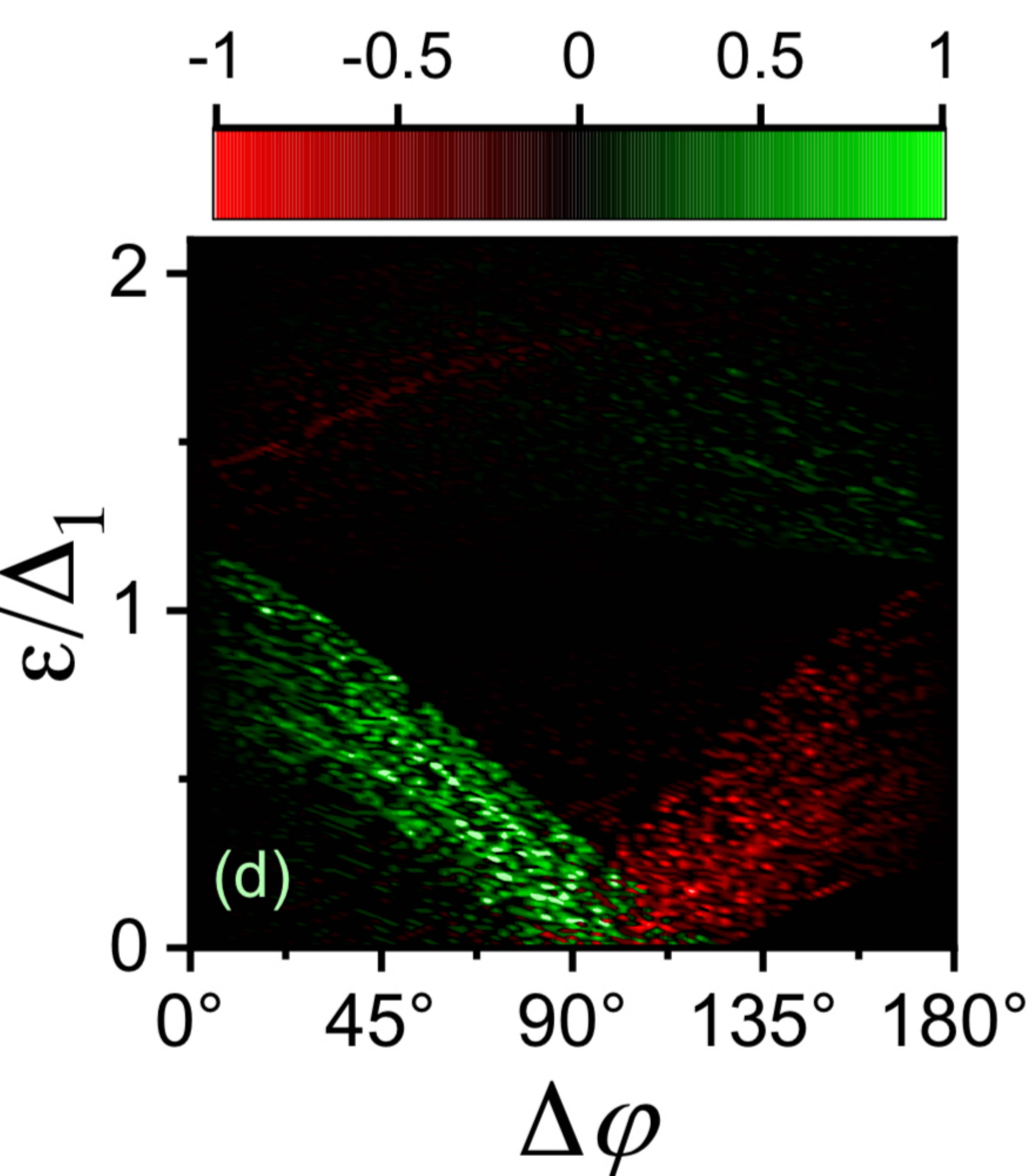} 
 }
 {
 \includegraphics[width=0.24\textwidth,scale=0.01]{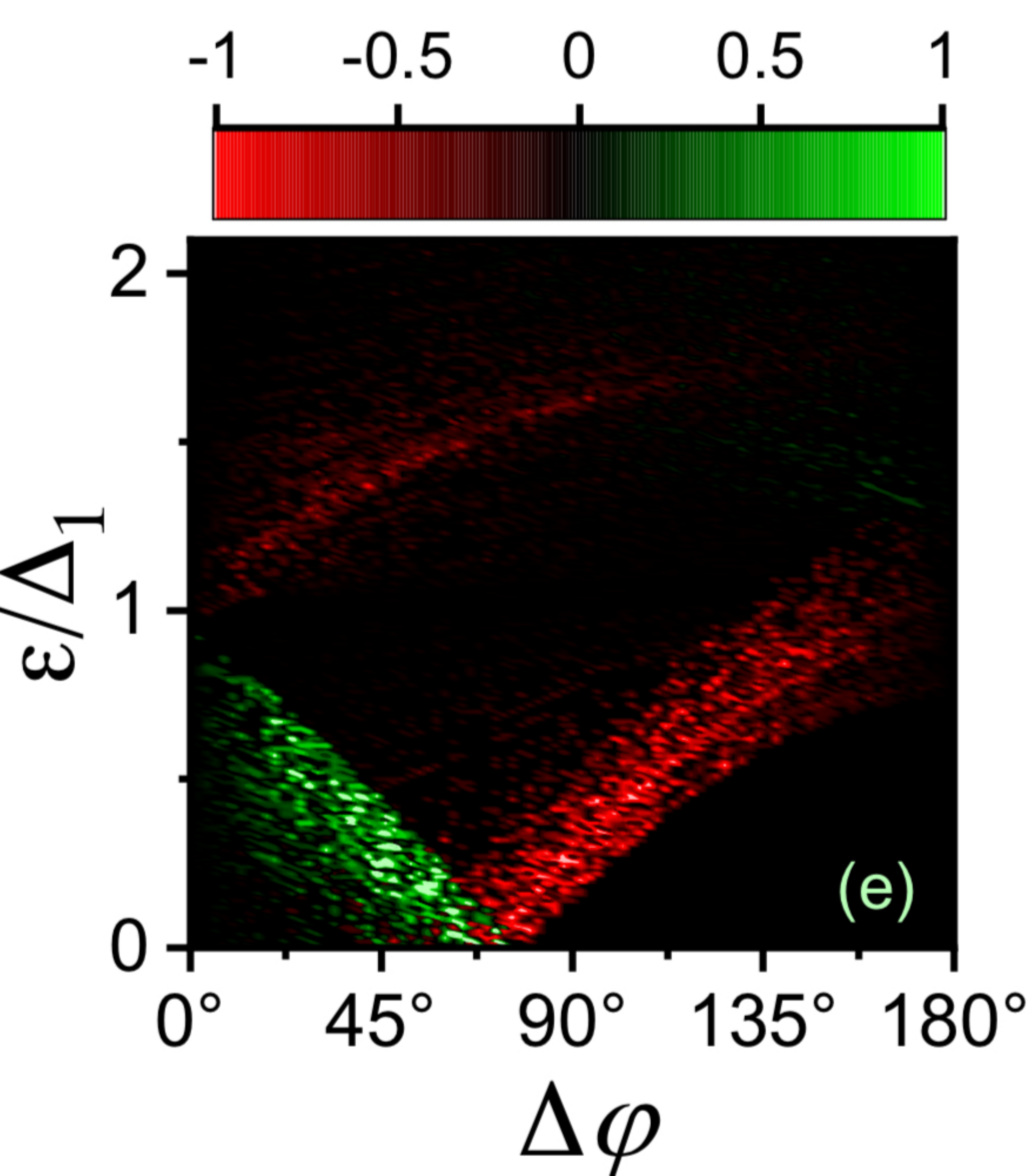} 
\includegraphics[width=0.24\textwidth,scale=0.01]{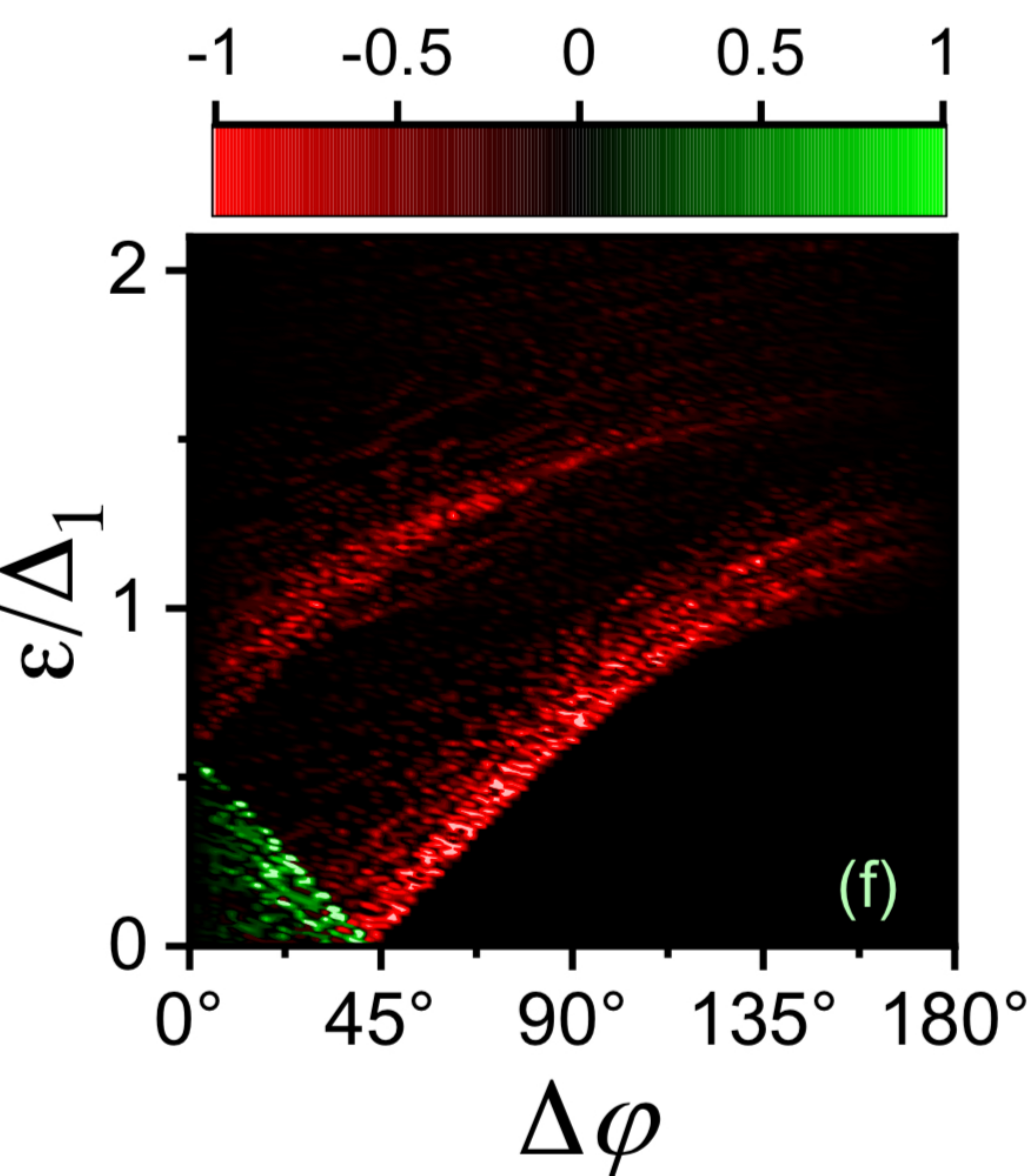} 
\includegraphics[width=0.24\textwidth,scale=0.01]{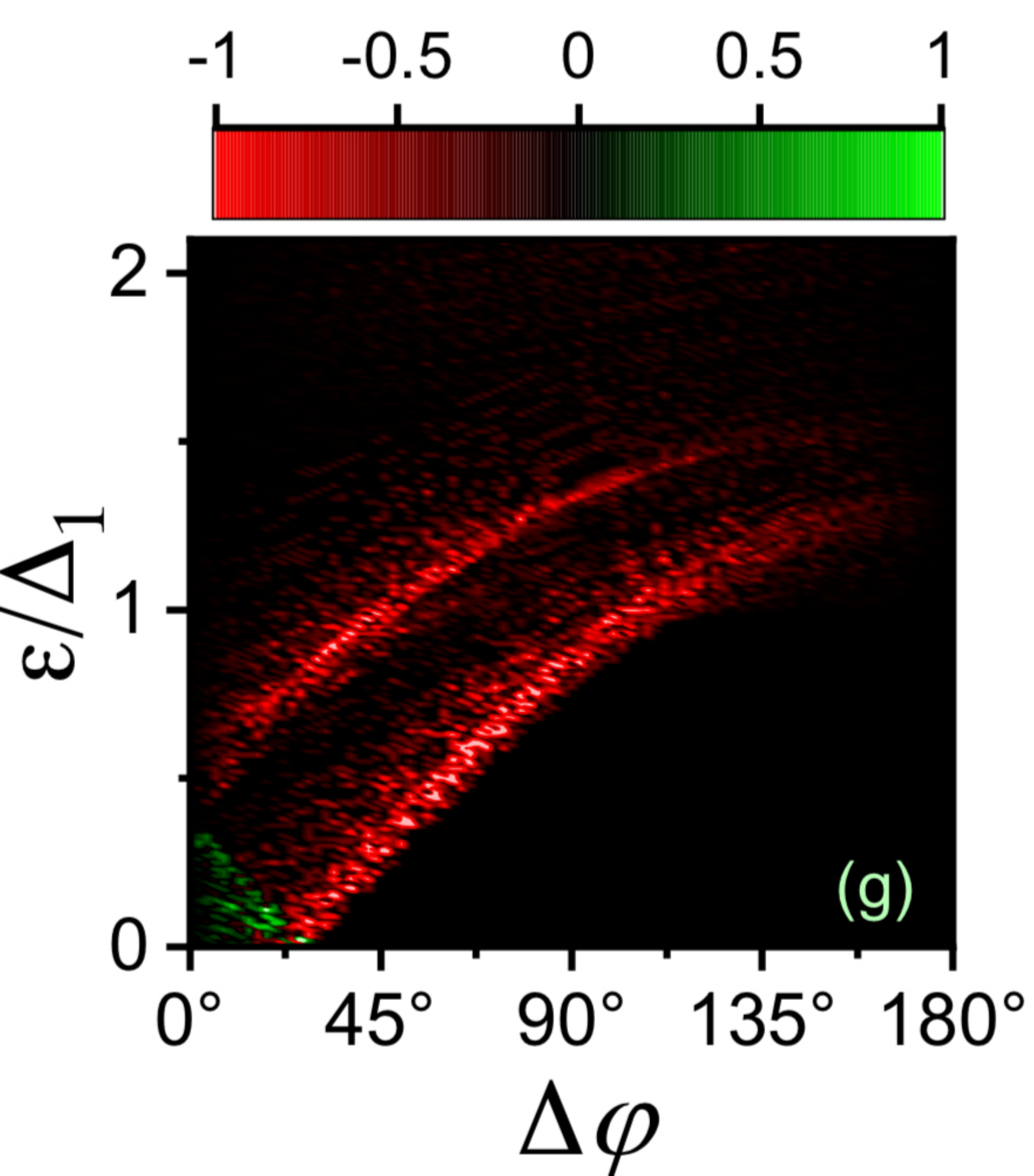} 
\includegraphics[width=0.24\textwidth,scale=0.01]{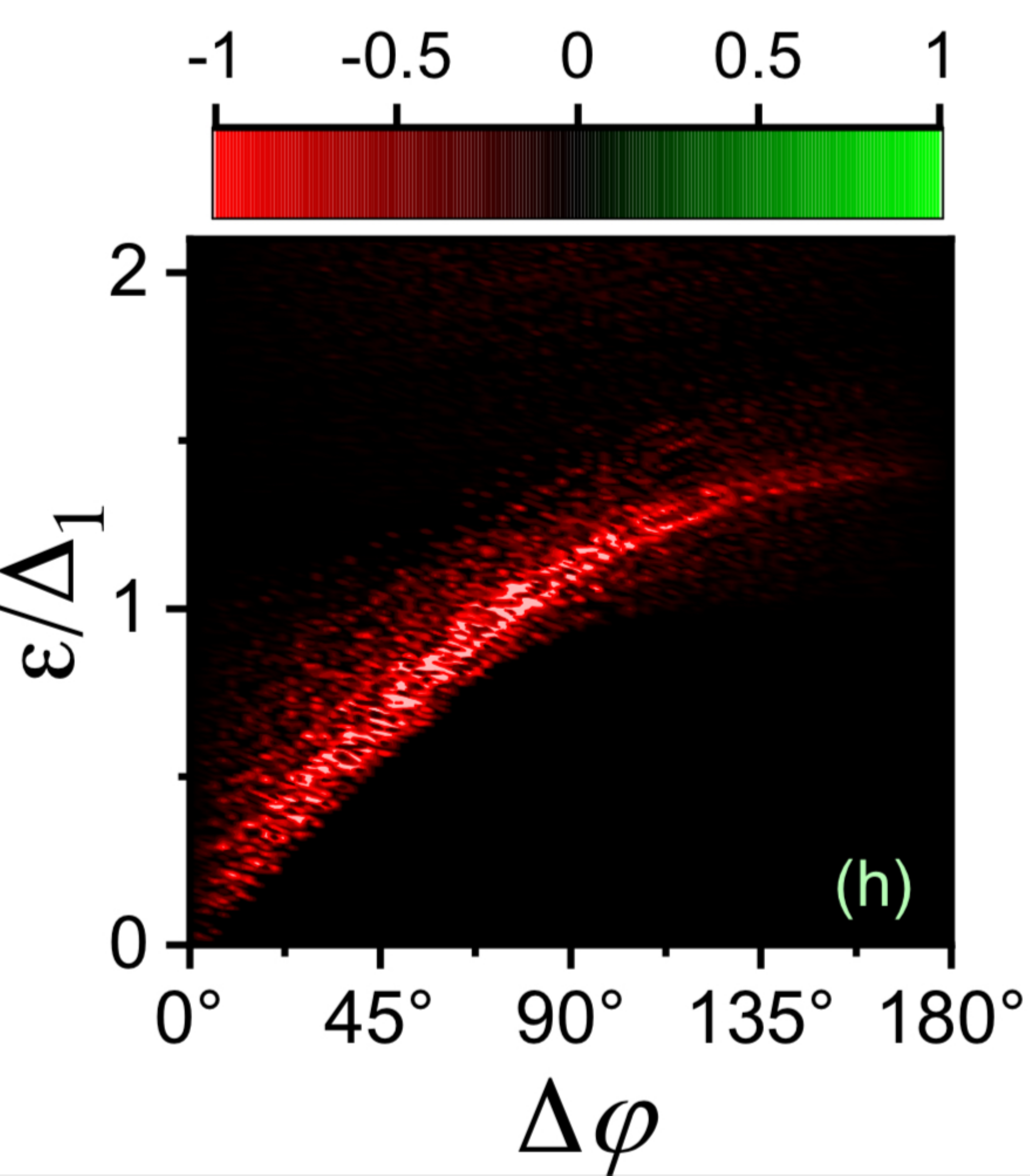}  
 }
\caption{(Color online). Total supercurrent density mappings as a function of phase difference $\Delta\varphi$ and 
energy $\varepsilon$ for an asymmetric Josephson junction with $\Delta_2/\Delta_1=5$. 
The Fermi level mismatch parameter is set to $\Lambda = 3$.
Panels (a)-(h) depict differing normalized exchange fields  $h/\Delta_1$
corresponding to $0,45,70,90,110,130,140$, and $160$, respectively.
 }
\label{fig2D}
\end{figure*}

\begin{figure*}[t!] 
\centering
\resizebox{0.98\textwidth}{!}{
\includegraphics{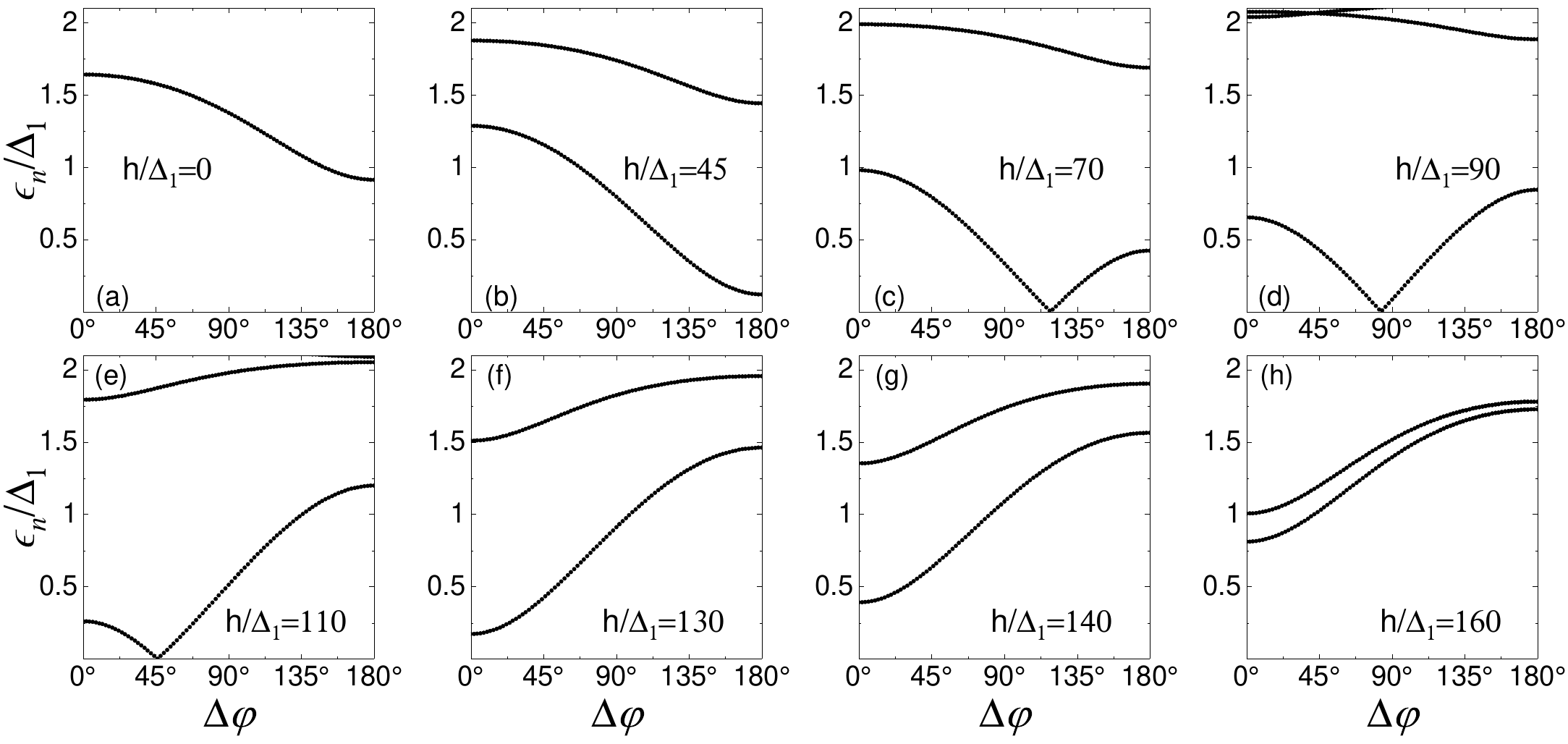}}
\caption{
(Color online).  The discrete energy spectrum for an asymmetric Josephson 
junction with $\Delta_2/\Delta_1=5$,
and Fermi level mismatch corresponding to $\Lambda=3$.
The normalized exchange fields considered correlate with Fig.~\ref{fig2D} above.
}
\label{ephase}
\end{figure*}

\section{Alternative wave-function approach: ballistic regime}\label{alt}
The method outlined in Sec.~\ref{theory-ballistic} provides an effective way
to solve Josephson junction systems with limited 
approximations in the ballistic regime.
There also exists wave-function 
approaches that
can provide exact solutions, one of which we outline below\cite{AlidoustWS1,AlidoustBP1,M.Alidoust2020,AlidoustBP2,AlidoustWS2}. 

To simulate the low-energy physics of  
heterostructures in the presence of a magnetization with arbitrary direction, 
$\mathbf{h}=(h_x,h_y,h_z)$, one employs an effective single-particle Hamiltonian:
\begin{equation}\label{Hamil_1}
H =\frac{1}{2}\int d\mathbf{p}~ \hat{\psi}^{\dag}(\mathbf{p}) \left[\frac{\mathbf{p}^2}{2m} +\bsigma\cdot  \mathbf{h}\right]\hat{\psi}(\mathbf{p}),
\end{equation}
where $\mathbf{p}=(p_x,p_y,p_z)$ is the momentum and $m$ is the effective mass of a charged particle. The associated field operator in spin space is given by $\hat{\psi}=(\psi_{\uparrow}, \psi_{\downarrow})^{\mathrm{T}}$ and $\bsigma=(\sigma_{x},\sigma_{y}, \sigma_{z})$ is a vector comprised of Pauli matrices. The spin-singlet superconductivity in the BCS scenario can be described by the following electron-electron amplitudes:
\begin{equation}
\Delta \langle \psi_\uparrow^\dag \psi_\downarrow^\dag \rangle + \text{H.c.}~.
\end{equation}  
Accounting for the electron-electron amplitudes in the BdG formalism, the low-energy Hamiltonian in spin-Nambu space reads:
\begin{equation}\label{Hamil}
{\cal H}(\mathbf{p}) = \left( \begin{array}{cc}
 H(\mathbf{p}) -\mu \hat{1}& \hat{\Delta} \\
 \hat{\Delta}^\dag & -H^\dag(-\mathbf{p}) +\mu \hat{1}
\end{array}\right),
\end{equation}
in which $\hat{\Delta}$ is the superconducting 
gap $\hat{\Delta}\equiv(\Delta,-\Delta)e^{i\varphi_{l,r}}$, and $\mu$ is the chemical potential. 
The field operators in the rotated spin-Nambu space are given by $\hat{\psi}=(\psi_{\uparrow}, \psi_{\downarrow}, \psi_{\downarrow}^{\dag}, -\psi_{\uparrow}^{\dag})^{\mathrm{T}}$. 
The continuity equation for charged carriers is written:
\begin{equation}
\begin{split}
\frac{\partial \rho_\text{c}}{\partial t}=\lim\limits_{\mathbf{r}\rightarrow \mathbf{r}'}\sum\limits_{\sigma\tau\sigma'\tau'}\frac{1}{i}\Big[ \psi^\dag_{\sigma\tau}(\mathbf{r}'){\cal H}_{\sigma\tau\sigma'\tau'}(\mathbf{r})\psi_{\sigma'\tau'}(\mathbf{r})\\-\psi^\dag_{\sigma\tau}(\mathbf{r}'){\cal H}_{\sigma\tau\sigma'\tau'}^\dag(\mathbf{r}')\psi_{\sigma'\tau'}(\mathbf{r})\Big],
\end{split}
\end{equation}
where ${\cal H}_{\sigma\tau\sigma'\tau'}$ is the component form of Eq.~(\ref{Hamil}) and $\sigma, \tau$ indices label the spin and particle-hole degrees of freedom, respectively. 
In a situation where no sink or source of charge is present, the time variation of charge density vanishes:
$\partial_t\rho_\text{c}\equiv~0$.
Accounting for the current conservation law, the charge current density can be expressed by:
\begin{equation}\label{crntbls}
\mathbf{ J} =\int \hspace{-.1cm} d\mathbf{r}\Big\{\hat{\psi}^\dagger(\mathbf{r}) \overrightarrow{{\cal H}}(\mathbf{r})\hat{\psi}(\mathbf{r})-
\hat{\psi}^\dagger(\mathbf{r}) \overleftarrow{{\cal H}}(\mathbf{r})\hat{\psi}(\mathbf{r}) \Big\},
\end{equation}
where the real-space Hamiltonian  ${\cal H}(\mathbf{r})$ is obtained by substituting 
$\mathbf{ p}\equiv~{-i m^{-1} (\partial_x,\partial_y,\partial_z)}$
 in Eq.~(\ref{Hamil}). 
The arrows indicate the specific wave functions that the Hamiltonian operates on. To obtain the supercurrent, one computes the current density perpendicular to the interfaces, in our geometry shown in Fig.~\ref{fig1}, 
$J_x$, and integrates over the junction cross section in the $y$ direction: 
$J(\Delta\varphi)=J_0\int_{-W/2}^{+W/2}  dy J_x(x,y,\Delta\varphi)$.  
Here $J_0=2e |\Delta| /\hbar$, and $e$ is the electron charge.
 Upon diagonalizing ${\cal H}(\textbf{p})$ in  Eq.~(\ref{Hamil}), one obtains 
 the electronic wave functions $\hat{\psi}_{l,m,r}(\textbf{p})$ within the 
  left ($l$), middle ($m$), and right ($r$) regions. 
  Next, the wave functions are matched at the left $\hat{\psi}_l$=$\hat{\psi}_m|_{x=0}$ and the right boundaries $\hat{\psi}_m$=$\hat{\psi}_r|_{x=d}$.
 The continuity conditions must also be satisfied:
  $(\partial_\textbf{p} {\cal H}_l)_{\textbf{r}}\hat{\psi}_l$=$(\partial_\textbf{p} {\cal H}_m)_{\textbf{r}}\hat{\psi}_m|_{x=0}$, 
  $(\partial_\textbf{p} {\cal H}_m)_{\textbf{r}}\hat{\psi}_m$=$(\partial_\textbf{p} {\cal H}_r)_{\textbf{r}}\hat{\psi}_r|_{x=d}$. 
  The index $\textbf{r}$ indicates a
   switch to real space after taking the derivatives in momentum space.
    It is important to note that we apply no simplifying assumptions and approximations to the wave functions in 
    the  numerical calculations. This however results in 
    highly complicated and lengthy expressions for the wave functions and supercurrent. 
    We therefore are only able to evaluate them numerically.

\section{energy dispersion and energy-resolved supercurrent density: ballistic regime}\label{e-resolved_ballistic}
When calculating the supercurrent via Eq.~(\ref{bdgcurrent}),
all positive energy states within an energy cutoff $\epsilon_c$
are summed over, with $\epsilon_c$ chosen sufficiently large
so that including any additional states has no distinguishable effects on the results.
It is important to note that 
when summing the quantum states for the supercurrent,
the summation implicitly includes an integration over
the continuum of states with transverse energy $\epsilon_\perp$ [see Eq.~(\ref{hoho})].
To isolate the supercurrent contribution at  supergap and subgap energies, 
it is beneficial to
extract the  supercurrent response as a function of the quasiparticle energy $\epsilon$.
This procedure 
involves  calculating the supercurrent  for each quasiparticle trajectory 
with associated energy  $\epsilon_\perp$. All trajectories
are then summed over to arrive at  the supercurrent
for a given energy.

This procedure gives the results shown in Fig.~\ref{fig2D_sns}, where the 
supercurrent is mapped out as a function of energy and phase difference.
For concreteness, we take the parameters used in Figs.~\ref{fig7}(a)-\ref{fig7}(d),
where a broad range of gap asymmetries were considered.
Note the emergence of the $J_2$ supercurrent in the current phase relations 
can be accounted for in Figs.~\ref{fig2D_sns}(b)-\ref{fig2D_sns}(d) where
the current carrying states get shifted upwards  into the supergap region and 
broaden with increased $\Delta_2/\Delta_1$.
This also leads to an amplification of the $J_2$ component, and
in turn the total supercurrent.

  \begin{figure*}
\centering
\includegraphics[width=1\textwidth]{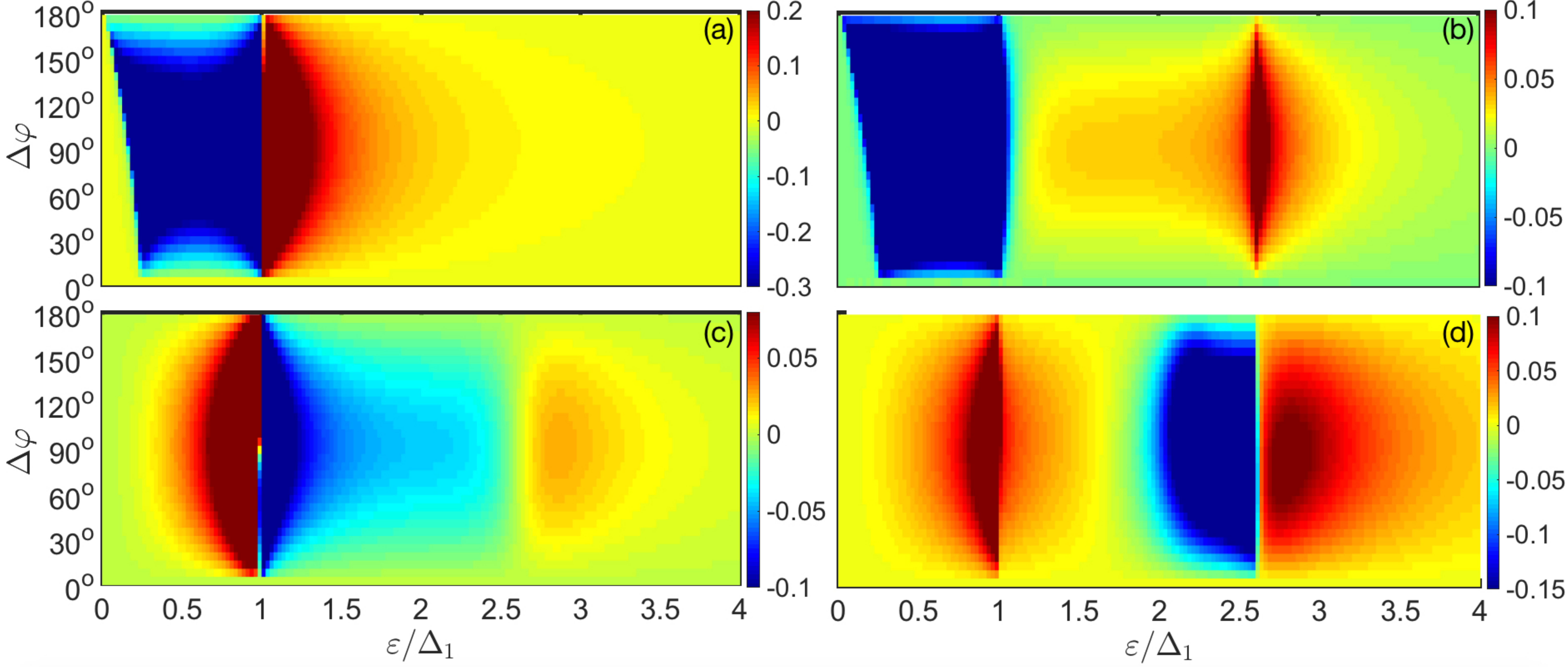} 
\caption{(Color online). Color map of the supercurrent density as a function of 
the normalized quasiparticle
energy $\varepsilon/\Delta_1$, and superconducting phase difference $\Delta\varphi$ in the diffusive regime. The junction parameter values are set in each panel to those of Fig.~\ref{I123_diff}, except now
we have  (a) \sns: $h=0$, $\Delta_2/\Delta_1=1$, (b) \sns: $h=0$, $\Delta_2/\Delta_1=2.6$, (c) \sfs: $h=2.6\Delta_1$, $\Delta_2/\Delta_1=1$, and (d) \sfs: $h=2.6\Delta_1$, $\Delta_2/\Delta_1=2.6$.}
\label{I_E_phi_diff}
\end{figure*}

The energy-resolved supercurrent for an asymmetric 
Josephson junction at fixed phase is shown
in Fig.~\ref{jspike}.
We take 
 a relative ratio of $\Delta_2/\Delta_1=5$,
 and  mismatch in Fermi levels corresponding to  $\Lambda=3$.
Figure~\ref{jspike}(a) corresponds to a nonmagnetic junction while Figs.~\ref{jspike}(b) and \ref{jspike}(c) describe a ferromagnetic 
junction with differing exchange field strengths, as shown. 
For consistency, these system parameters correlate with
 Fig.~\ref{fig4}(g).
 To correlate with the critical current components, from Fig.~\ref{fig4}(g),
 it is seen that
 for $h/\Delta_1=0$  the supergap component $J_2$ dominates, while 
for  $h/\Delta_1=90$, the subgap $J_1$ component does.
The intermediate case of $h/\Delta_1=45$ has the subgap $J_1$ and 
 supergap $J_2$ components  contributing nearly equally to the 
 critical current.
 This behavior is 
 accounted for in the energy dependence  of the supercurrent 
 presented  in
 Figs.~\ref{jspike}(a)-\ref{jspike}(c), where we take  $\Delta\varphi = \Delta\varphi_c$
 in each case to ensure that it gives the critical current
shown in 
Fig.~\ref{fig4}(g).
 The shift in current carrying states is evident as the exchange energy increases,
 until eventually reversing direction for $h/\Delta_1=90$ where the subgap bound states
 dominate.

To give a more comprehensive  view of the energy-resolved supercurrent we present in 
Fig.~\ref{fig2D}, the normalized supercurrent as a function of
the normalized energy $\epsilon/\Delta_1$ and phase difference $\Delta\varphi$.
Eight different exchange fields are considered, and the system parameters again
coincide with the critical current study in Fig.~\ref{fig4}(g).
As seen in Figs.~\ref{fig2D}(a)-\ref{fig2D}(h), the supercurrent profiles exhibit varying 
amounts of mixtures between the subgap and supergap states.
As the exchange field increases, the supercurrent flow
evolves to contain states that have oppositely directed flow in both the subgap
($\epsilon \leq \Delta_1$)
and supergap ($\Delta_1\leq\epsilon \leq \Delta_2$) regions.
Remarkably, increasing the 
exchange field can increase the
supercurrent,
including   the half-metallic limit [Fig.~\ref{fig2D}(h)], 
where the 
 supercurrent can exceed the supercurrent  found in Fig.~\ref{fig2D}(a) for the
nonmagnetic case.
This can have important consequences for devices that utilize the
spin degree of freedom in  Josephson junction systems.
Note that the various admixtures of subgap and supergap supercurrents
exhibited here can be directly correlated with the critical current signatures found 
in Fig.~\ref{fig4}(g).

To delve further into the phase dependence of the supercurrent flow, we
next present in Fig.~\ref{ephase} the  quasiparticle energy spectra $\epsilon_n$ for
each of  the cases shown in Fig.~\ref{fig2D}.
We consider the $\epsilon_\perp = 0$ mode, noting that other transverse modes have similar behavior,
and
the cumulative effect serves to only broaden the overall supercurrent profile.
From the diagrams, it is clear that  the current vanishes at
$\Delta\varphi=0^\circ$ and $\Delta\varphi=180^\circ$, where $\partial \epsilon_n/\partial (\Delta\varphi) =0$.
The cusps in the energy dispersion 
are consistent with Fig.~\ref{fig2D}, where the current 
at certain energies becomes reversed.
We also find that as the exchange field $h$ increases, the additional branches of the energy dispersion
which 
emerge  increase in separation before coalescing  at  high exchange fields.
Thus, although the supercurrent is a cumulation of quasiparticle amplitudes and energies, 
the energy spectrum alone gives valuable insight into the transport properties of asymmetric 
Josephson junctions.

\section{energy-resolved supercurrent density: diffusive regime}\label{e-resolved_diff}
When impurity scattering dominates, we turn to the diffusive regime. In Fig.~\ref{I_E_phi_diff},
we show the supercurrent response as a function of macroscopic phase difference $\Delta \varphi$
and normalized energy $\epsilon/\Delta_1$. In Figs.~\ref{I_E_phi_diff}(a) and \ref{I_E_phi_diff}(b), a nonmagnetic junction 
$h=0$ is considered. 
In Fig.~\ref{I_E_phi_diff}(a), the junction is symmetric $\Delta_2/\Delta_1=1$, while in Fig.~\ref{I_E_phi_diff}(b), an asymmetric junction with 
$\Delta_2/\Delta_1=2.6$ is shown.
The bottom set of panels corresponds to a weak ferromagnet junction with $h/\Delta_1 = 2.6$.
For the nonmagnetic case, Figs.~\ref{I_E_phi_diff}(a) and \ref{I_E_phi_diff}(b) show how the supercurrent
with energies above the gap $\Delta_1$ get shifted by an amount corresponding to the gap asymmetry $\Delta_2/\Delta_1 = 2.6$.
In the bottom set of panels, the exchange field $h=2.6 \Delta_1$ introduces an additional energy scale 
that redistributes the supercurrent response. In the symmetric case Fig.~\ref{I_E_phi_diff}(c)
demonstrates how the presence of magnetism induces a supercurrent reversal for energies centered 
around the gap $\Delta_1$. When the junction becomes asymmetric,
the mutual effects of the exchange field and gap asymmetry lead to
 an enhancement of the supercurrent density at larger energies  
 around $\Delta_2$ [Fig.~\ref{I_E_phi_diff}(d)].

\end{document}